\pdfoutput=1

\documentclass[11pt]{article}

\usepackage[preprint]{acl}

\usepackage{times}
\usepackage{latexsym}

\usepackage[T1]{fontenc}

\usepackage[utf8]{inputenc}

\usepackage{microtype}

\usepackage{inconsolata}

\usepackage{graphicx}

\usepackage{microtype}
\usepackage{inconsolata}

\usepackage{graphicx}
\usepackage{tabularx}
\usepackage{adjustbox}
\usepackage{booktabs}
\usepackage{multirow}
\usepackage{amsmath}
\usepackage{cuted}
\usepackage{comment}
\usepackage{lipsum}
\usepackage{colortbl}
\usepackage{xcolor}
\usepackage{enumitem}
\usepackage{bbding}
\usepackage{subcaption}
\usepackage{amssymb}
\usepackage{amsfonts}

\usepackage{xspace}

\definecolor{cmzhao}{rgb}{0.1, 0.8, 0.1}

\definecolor{cmwu}{rgb}{0.8, 0.1, 0.1}

\def\mname{DeepResonance\xspace}

\title{DeepResonance: Enhancing Multimodal Music Understanding \\ via Music-centric Multi-way Instruction Tuning}

\author{
    Zhuoyuan Mao$^1$\thanks{\ \ Currently at Tencent. Work done while at Sony Group Corporation.} \hspace{1em} Mengjie Zhao$^1$\thanks{\ \ Currently at SB Intuitions. Work done while at Sony Group Corporation.} \hspace{1em} Qiyu Wu$^1$ \\ \textbf{Hiromi Wakaki}$^1$ \hspace{1em} \textbf{Yuki Mitsufuji}$^{1,2}$ \\
    $^1$Sony Group Corporation \hspace{1em} $^2$Sony AI \\
    \texttt{kevinmzy@gmail.com, mengjie@fastmail.com} \\ \texttt{\{qiyu.wu, hiromi.wakaki, yuhki.mitsufuji\}@sony.com}
}

\begin{document}
\maketitle
\begin{abstract}


Recent advancements in music large language models (LLMs) have significantly improved music understanding tasks, which involve the model’s ability to analyze and interpret various musical elements. These improvements primarily focused on integrating both music and text inputs. However, the potential of incorporating additional modalities such as images, videos and textual music features to enhance music understanding remains unexplored. To bridge this gap, we propose DeepResonance, a multimodal music understanding LLM fine-tuned via multi-way instruction tuning with multi-way aligned music, text, image, and video data. To this end, we construct Music4way-MI2T, Music4way-MV2T, and Music4way-Any2T, three 4-way training and evaluation datasets designed to enable DeepResonance to integrate both visual and textual music feature content. We also introduce multi-sampled ImageBind embeddings and a pre-LLM fusion Transformer to enhance modality fusion prior to input into text LLMs, tailoring for multi-way instruction tuning. Our model achieves state-of-the-art performances across six music understanding tasks, highlighting the benefits of the auxiliary modalities and the structural superiority of DeepResonance. We open-source the codes, models and datasets we constructed: \url{https://github.com/sony/DeepResonance}.




\end{abstract}

\begin{figure}[t]
    \centering
    \includegraphics[width=0.95\linewidth]{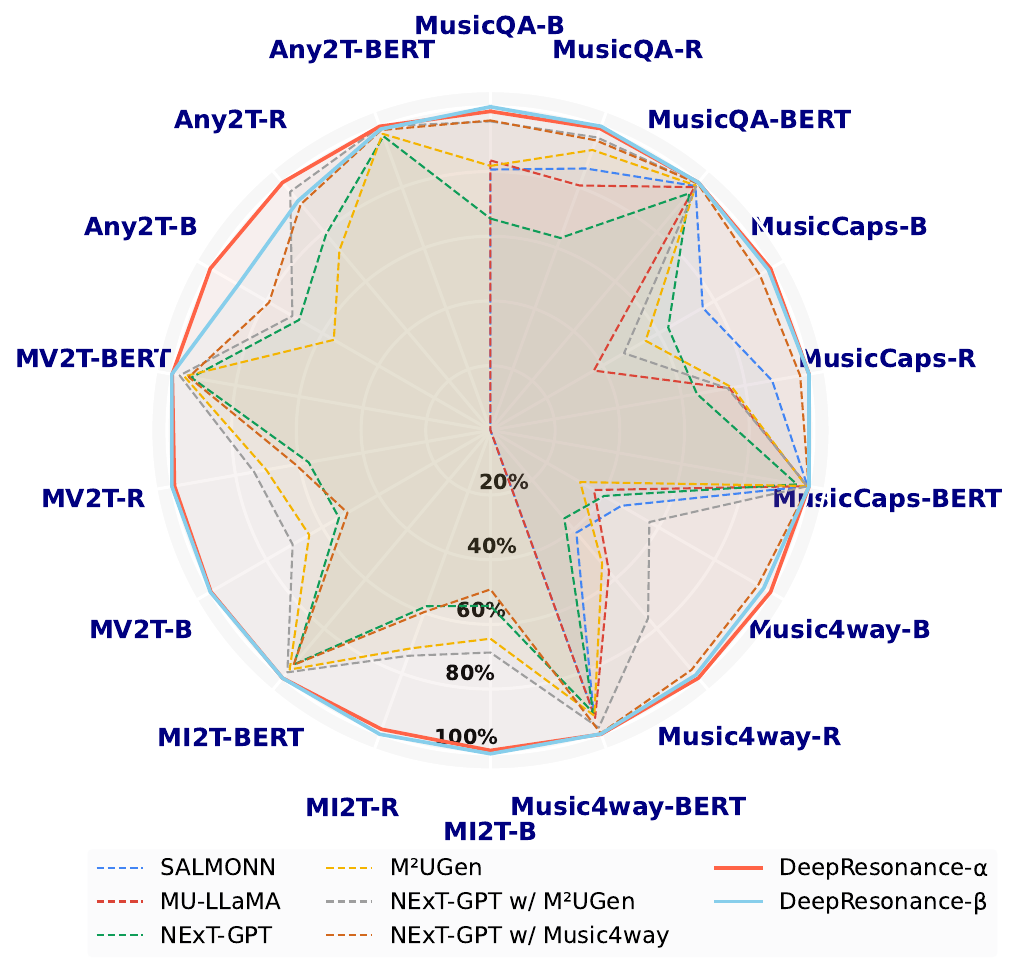}
    \caption{Performance overview of DeepResonance and related models. ``Music4way,'' ``MI2T,'' ``MV2T,'' and ``Any2T'' refer to the Music4way-MusicCaps, Music4way-MI2T, Music4way-MV2T, and Music4way-Any2T datasets. The metrics ``B,'' ``R,'' and ``BERT'' represent BLEU-1, ROUGE-L, and BERTScore. All values have been normalized based on maximum.} 
    \label{fig:radar}
\end{figure}

\section{Introduction}
\begin{quote}
    \textit{``Music gives a soul to the universe, wings to the mind, flight to the imagination, and life to everything.''}  
    \begin{flushright}
        --- Plato
    \end{flushright}
\end{quote}

Different modalities are often interwoven. In the context of music, humans typically experience music alongside complementary textual and visual signals, such as lyrics, the composition of a piece, live performances, or the arrangement of instruments in a band. These additional modalities significantly influence how humans perceive and understand music. Therefore, it is reasonable to assume that training a model to incorporate other modalities—such as images, videos, and textual music features can enhance its performance on music understanding tasks~\cite{DBLP:conf/ijcnn/MancoBQF21,DBLP:conf/icml/GardnerDSB24,DBLP:journals/corr/abs-2301-11325,DBLP:conf/icassp/LiuHSS24}.

In this work, we introduce the concept of multimodal music understanding, where multiple modality signals are leveraged to enhance the perception of music. We implement this concept through multi-way instruction tuning, inspired by code-switched~\cite{DBLP:conf/iclr/SongZQXCWL22} and multi-way multilingual translation~\cite{DBLP:journals/jmlr/FanBSMEGBCWCGBL21} that extend translation models beyond bilingual pairings. For multimodal music understanding, we establish relationships that go beyond the conventional pairing of music and text modalities commonly seen in existing music large language models (LLMs)~\cite{DBLP:conf/ismir/DohCLN23,DBLP:conf/icassp/LiuHSS24,DBLP:journals/corr/abs-2311-11255,DBLP:conf/icml/GardnerDSB24,DBLP:journals/corr/abs-2410-15573}.

We present \textbf{DeepResonance}, a multimodal music understanding LLM trained on music-centric multi-way data that integrates the music, text, image, and video modalities. Specifically, we construct two 4-way training datasets, \textbf{Music4way-MI2T} and \textbf{Music4way-MV2T}, where the source data comprises music, images/videos, and textual descriptions and instructions. The target data includes textual descriptions enriched with multimodal information, such as music, images/videos, and low-level music features, including tempo, chords, key, and downbeats. Using these datasets, we develop DeepResonance based on the NExT-GPT~\cite{DBLP:conf/icml/Wu0Q0C24} architecture. To enhance multimodal music understanding tasks, we propose two key modifications to the backbone model. The first involves \textbf{multi-sampled ImageBind embeddings}, which are designed to retain richer information of music, image, and video modalities from the ImageBind encoders~\cite{DBLP:conf/cvpr/GirdharELSAJM23}, thereby fostering deeper interaction with the music modality and improving multimodal music understanding. The second is a \textbf{pre-LLM fusion Transformer}, a module aimed at pre-adapting different modalities to each other before feeding them into the text LLM module. This component is particularly effective in simultaneously handling multimodal inputs of music, text, images, and videos. These innovations collectively advance the model's ability to integrate and process diverse multimodal signals for enhanced music understanding.


We evaluate DeepResonance on three conventional music understanding tasks (i.e., music + text (instruction)$\xrightarrow{}$ text) and three multimodal music understanding tasks (i.e., music + image/video + text (instruction) $\xrightarrow{}$ text (multimodal-enriched)). The former includes two existing benchmarks, MusicQA~\cite{DBLP:conf/icassp/LiuHSS24} and MusicCaps~\cite{DBLP:journals/corr/abs-2301-11325}, along with our constructed Music4way-MusicCaps, which cover captioning and question-answering tasks for music understanding. The latter evaluation for multimodal music understanding uses the test splits of Music4way-MI2T and Music4way-MV2T, as well as the newly introduced \textbf{Music4way-Any2T} to assess the model's robustness.

As shown in Fig.~\ref{fig:radar}, DeepResonance models with different configurations ($\alpha$ and $\beta$) consistently outperform related models across all six downstream tasks in supervised settings, demonstrating the effectiveness of our proposed multi-way datasets and model architecture components for music understanding tasks. Additionally, we conduct zero-shot evaluations to assess the model's generalization to unseen datasets and perform ablation studies to evaluate the impact of each proposed component in different downstream task settings. The contributions of this work can be summarized as follows:
\begin{itemize}
    \item We introduce the Music4way-MI2T, Music4way-MV2T, and Music4way-Any2T datasets, enabling music, text, image, and video integration for music understanding.
    \item We propose multi-sampled ImageBind embeddings and a pre-LLM fusion Transformer to enhance multimodal fusion for music LLMs.
    \item Our DeepResonance outperforms existing music LLMs across six downstream tasks, showing the effectiveness of our proposed datasets and models, which will be open-sourced.
\end{itemize}

\begin{figure*}[t]
    \centering
    \includegraphics[width=0.97\linewidth]{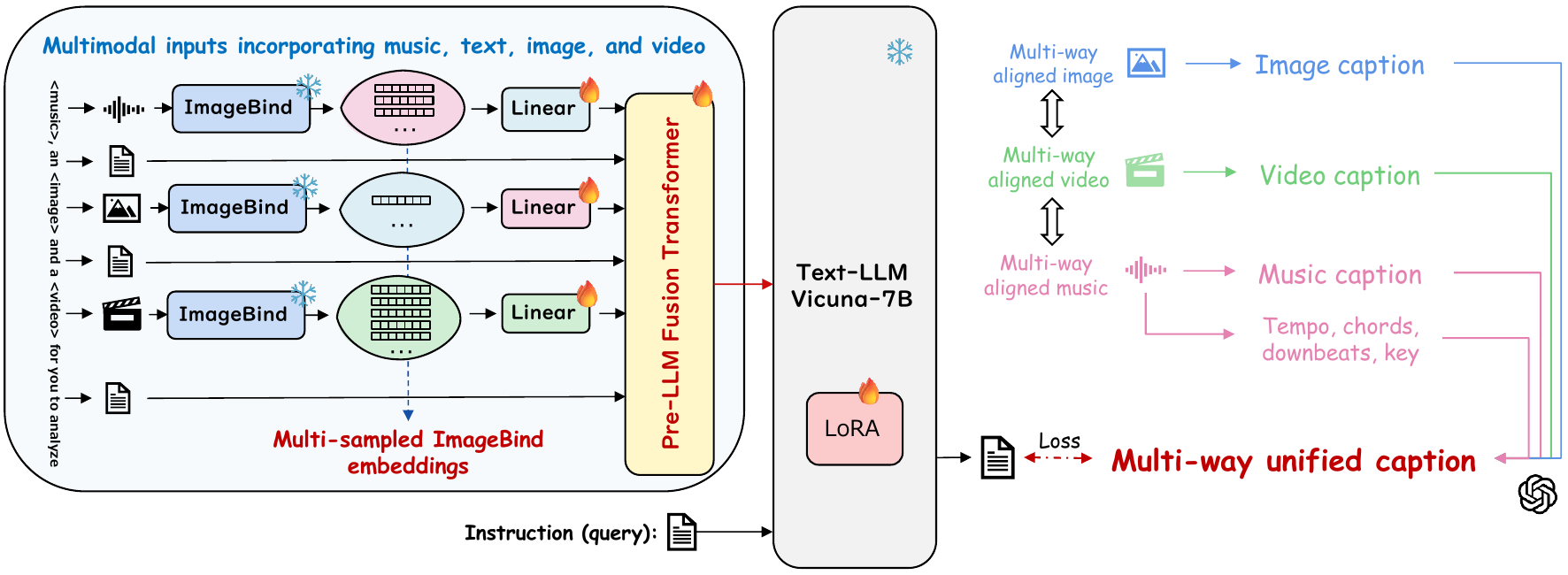}
    \caption{\textbf{Overview of DeepResonance.} It integrates the ``multi-way unified caption'' as a target, along with ``multi-sampled ImageBind embeddings'' and a ``pre-LLM fusion Transformer'' as novel architectural components.}
    \label{fig:model}
\end{figure*}

\section{Related Work}
\noindent \textbf{Music Understanding}
is an emerging topic that builds upon the foundational research efforts in music information retrieval (MIR), which traditionally focused on low-level music feature recognition tasks, such as identifying tempo, chords, keys, and instruments~\cite{DBLP:conf/ecir/FaraldoGJH16,DBLP:conf/ismir/PauwelsOGS19,DBLP:conf/ismir/GururaniSL19,DBLP:journals/tismir/SchreiberUM20}. Early work in this area was centered on basic tagging tasks, such as determining the genre or version of a piece of music~\cite{DBLP:conf/ismir/Tzanetakis01,DBLP:conf/icassp/WonONGS21,DBLP:journals/spm/YesilerDBTS21}. Over time, the focus shifted to high-level understanding tasks that require a more comprehensive interpretation of the content, sentiment, and insights conveyed by music. These tasks include captioning, reasoning, question answering, and tool using~\cite{DBLP:conf/ijcnn/MancoBQF21,DBLP:conf/icml/GardnerDSB24,DBLP:journals/corr/abs-2301-11325,DBLP:conf/icassp/LiuHSS24,deng-etal-2024-musilingo,DBLP:journals/corr/abs-2410-15573}.


\noindent \textbf{Multimodal Instruction Tuning and Music Foundation Models:}
Recently, multimodal pre-training has successfully bridged image, audio, and video modalities to text LLMs~\cite{DBLP:conf/iclr/TangYSC000M024,DBLP:conf/icml/Wu0Q0C24,DBLP:conf/iclr/0001LLKG24,DBLP:conf/cvpr/TangYKLZB24} through multimodal instruction tuning~\cite{DBLP:conf/nips/LiuLWL23a,DBLP:journals/corr/abs-2307-08581} or universal multimodal embedding space encoders~\cite{DBLP:conf/cvpr/GirdharELSAJM23,DBLP:conf/iclr/ZhuLNYCWPJZLZ0024}. However, few studies have focused on the music modality. MU-LLaMA~\cite{DBLP:conf/icassp/LiuHSS24} was among the first to instruction-tune LLaMA models~\cite{DBLP:journals/corr/abs-2302-13971} for the music domain, while LLark~\cite{DBLP:conf/icml/GardnerDSB24} extended music LLMs to support a wide range of tasks, including captioning, reasoning, and low-level music feature recognition. M$^2$UGen~\cite{DBLP:journals/corr/abs-2311-11255} introduced music generation modules built on MU-LLaMA, leveraging newly constructed music-centric datasets for instruction fine-tuning. OpenMU~\cite{DBLP:journals/corr/abs-2410-15573} unified existing music understanding datasets, curating a comprehensive benchmark for evaluating music + text $\xrightarrow{}$ text tasks. Other models, such as MusCaps~\cite{DBLP:conf/ijcnn/MancoBQF21}, LP-MusicCaps~\cite{DBLP:conf/ismir/DohCLN23}, and MusiLingo~\cite{deng-etal-2024-musilingo}, were designed for task-specific purposes.

Our work differs from the previous studies on music understanding and music foundation models in three aspects: (1) We are the first to apply instruction-tuning on music foundation models using multi-way data that integrates multiple modalities, shifting the paradigm from the traditional music + text $\xrightarrow{}$ text approach to the music + image/video + text $\xrightarrow{}$ text (multimodal-enriched) framework. (2) We are the first to access the generalization of music LLMs to multi-way multimodal inputs using our newly curated Music4way-Any2T dataset and zero-shot evaluation on out-of-domain benchmarks. (3) We propose multi-sampled ImageBind embeddings and pre-LLM fusion Transformer that significantly impact multimodal music understanding tasks, delivering state-of-the-art music LLMs optimized for different downstream tasks.

\begin{figure*}[t]
    \centering
    \includegraphics[width=0.98\linewidth]{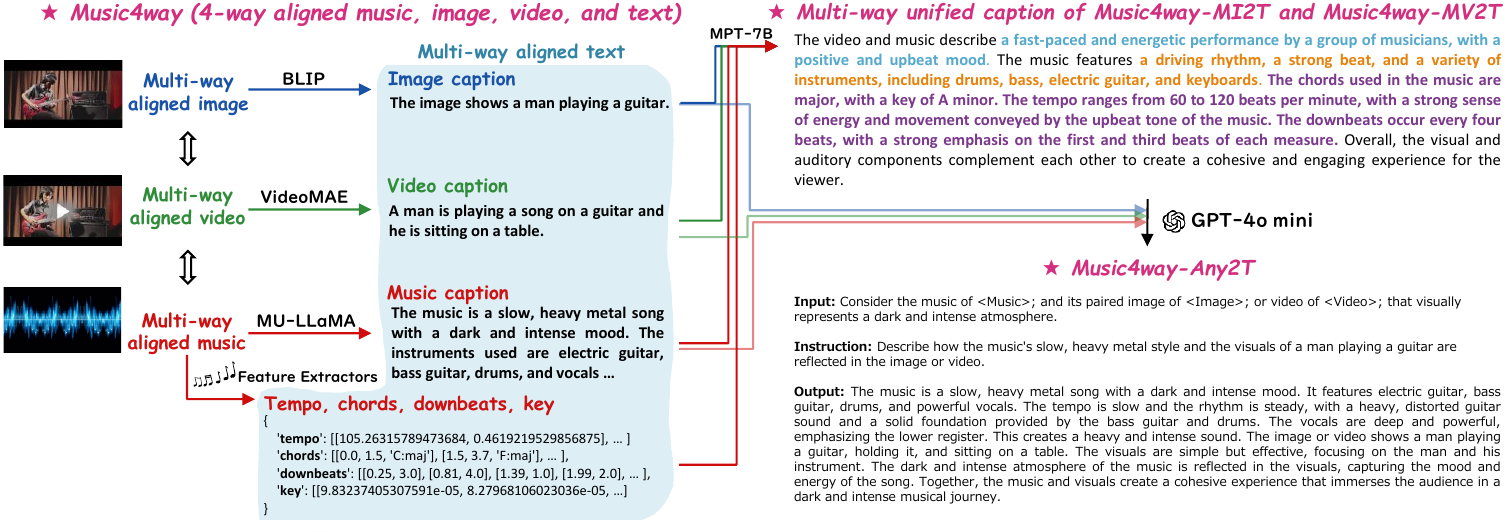}
    \caption{\textbf{Multi-way Instruction Tuning Data Construction.} Based on AudioSet, we construct Music4way (M+T$\xrightarrow{}$T, I+T$\xrightarrow{}$T, V+T$\xrightarrow{}$T), Music4way-MI2T (M+I+T$\xrightarrow{}$T), Music4way-MV2T (M+V+T$\xrightarrow{}$T), and Music4way-Any2T (M+I/V+T$\xrightarrow{}$T) for instruction tuning and evaluation. (M: music; I: image; V: video; T: text)}
    \label{fig:data}
\end{figure*}

\section{Multi-way Instruction Tuning}
We focus on the integration of multi-way information into each data for music understanding tasks, which conventionally contains only music and text. 



\subsection{Music-centric Multi-way Datasets Construction}
\label{sec:4way}
M$^2$UGen~\cite{DBLP:journals/corr/abs-2311-11255} pioneered multimodal dataset construction for instruction-tuning music LLMs, creating music-centric pairs with text, images, videos, and other music for captioning, editing, and generation tasks. Drawing inspiration from studies on code-switched and multi-way translation~\cite{DBLP:conf/iclr/SongZQXCWL22,DBLP:journals/jmlr/FanBSMEGBCWCGBL21}, which break away from English-centric bilingual pairing relationships, we propose extending traditional music LLM fine-tuning by incorporating multi-way relationships among music and other modalities, including image, video, and text. Building upon the music-centric paired dataset construction pipeline of M$^2$UGen, we expand it with multi-way relationships to create multimodal-enriched training data.


\noindent \textbf{Music4way:} Building upon the AudioSet~\cite{DBLP:conf/icassp/GemmekeEFJLMPR17} music clips filtered by M$^2$UGen, we curate a multi-way aligned dataset, comprising a total of 172.57 hours of music.\footnote{Note that in M$^2$UGen, each music clip is paired with only one other modality during training.} Each video-music pair from AudioSet is processed as shown in Fig.~\ref{fig:data} (left). First, we randomly extract a video frame to serve as the aligned image for the video and music. Next, BLIP~\cite{DBLP:conf/icml/0001LXH22}, VideoMAE~\cite{DBLP:conf/nips/TongS0022}, and MU-LLaMA~\cite{DBLP:conf/icassp/LiuHSS24} are used to generate captions for the image, video, and music, respectively. Following LLark~\cite{DBLP:conf/icml/GardnerDSB24}, we employ \texttt{madmom}~\cite{DBLP:conf/mm/BockKSKW16} to extract low-level music features, including tempo, chords, downbeats, and key, representing these features in textual form. This process creates a 4-way alignment for each music-video pair, establishing relationships between music, image, video, and text. The text modality conveys information through captions or feature values derived from the other three modalities. The resulting Music4way dataset includes 59,128 training samples (164.24 hours) and 3,000 evaluation samples (8.33 hours).

\subsection{Multi-way Instruction Tuning Data Construction}
\label{sec:data}
Building on the constructed Music4way dataset, we design instruction tuning data tailored for training and evaluating music LLMs. As fine-tuning an LLM requires data to be formatted strictly as source-target pairs, we develop two pipelines to transform the multi-way relationships among the four modalities in Music4way into source-target pairs, which ensure seamless integration of multimodal information for LLM instruction tuning.


\noindent \textbf{Music4way-MI2T and Music4way-MV2T:}
Inspired by VAST~\cite{DBLP:conf/nips/ChenLWZSZL23}, which established multimodal connections between vision, audio, and subtitles to generate omni-modal captions for improving visual-text tasks, we extend this approach to the music domain, exploring its potential benefits for music understanding tasks. As shown in Fig.~\ref{fig:data}, we prompt the open-source MPT-7B~\cite{MosaicML2023Introducing} model to combine captions from images, videos, and music, as well as textual music features, creating a multi-way unified caption. These captions are then paired with music and image/video inputs to form the source-target pairs required for fine-tuning LLMs. For the task of music + image to multi-way unified caption, we construct the Music4way-MI2T dataset. Similarly, for music + video inputs, we construct the Music4way-MV2T dataset. These datasets are designed to train the music understanding model to infer visual content and musical details (e.g., instruments, sentiments, and low-level features) directly from raw music and image/video files. The templates we utilized for prompting MPT-7B to generate the multi-way unified captions are provided in the Appx.~\ref{appx:mimv2t}. As shown in Fig.~\ref{fig:model}, these datasets are then used to fine-tune music understanding models, with inputs comprising music and image/video, a fixed instruction template (see Appx.~\ref{appx:mimv2t}), and outputs of multi-way unified captions. Additionally, we create corresponding test sets from the test split of Music4way to benchmark performance on multimodal music understanding tasks. (See Sec.~\ref{sec:mmmu})


\noindent \textbf{Music4way-Any2T:} To evaluate the robustness of multimodal music LLMs to diverse text inputs and queries, we introduce the Music4way-Any2T dataset. As shown in Fig.~\ref{fig:data}, this dataset features flexible inputs, allowing each modality to appear in any position. It includes instructions and outputs presented as diverse question-answer pairs, covering various aspects (e.g., visual or musical content) of the multi-way aligned data. To construct structured data with input, instruction, and output fields, we use GPT-4o mini\footnote{\url{https://platform.openai.com/docs/models}}, prompted with all textual information from the Music4way dataset alongside the multi-way unified caption. Detailed prompting templates and data examples can be found in the Appx.~\ref{appx:any2t}. The Music4way-Any2T dataset is used to benchmark the robustness and generalization capabilities of multimodal music understanding models, with results presented in Sec.~\ref{sec:mmmu}.


\section{Model Architecture Tailored for Multimodal Music Understanding}
\label{sec:model}
In this section, we introduce \textbf{DeepResonance}, our model designed to adapt general any-to-text LLMs for multimodal music understanding tasks, leveraging the multi-way datasets introduced in Sec.~\ref{sec:data}. We construct DeepResonance based on NExT-GPT~\cite{DBLP:conf/icml/Wu0Q0C24}, a general any-to-any multimodal LLM. This framework integrates an ImageBind encoder~\cite{DBLP:conf/cvpr/GirdharELSAJM23} to process inputs from 
the image, video, and audio modalities, a Vicuna-7B~\cite{vicuna2023} model as the LLM backbone, and linear adaptors to bridge ImageBind to the Vicuna model. The vanilla version of \mname, like NExT-GPT, models the text sequence generation task as follows:
\begin{align}
    \mathcal{P}(w_{n}|\mathbb{X}_m, \mathbb{X}_v, \mathbb{X}_i, \mathbb{X}_t, \mathbb{Q}, \mathbb{W}_{1:n-1}) = \nonumber \\
    \mathcal{LLM}(\mathcal{A}_m(\mathbf{e}_m), \mathcal{A}_v(\mathbf{e}_v), \mathcal{A}_i(\mathbf{e}_i), \nonumber \\
    \{\mathbf{e}_t\}, \{\mathbf{e}_q\}, \{\mathbf{e}_w\}_{1:n-1})
    \label{eq:nextgpt}
\end{align}
where $\mathbb{W} = \{w_1, w_2, \dots, w_n\}$ represents the text sequence to be generated, and $\mathbb{X}_m$, $\mathbb{X}_v$, $\mathbb{X}_i$ and $\mathbb{X}_t$ denote the patch-level multimodal and text tokens for music, video, image, and text, respectively. $\mathbb{Q}$ indicates the query (i.e., instruction) input to the model. $\mathcal{LLM}$ and $\mathcal{A}_*$ ($* \in \{m, v, i\}$) represent the Vicuna-7B LLM and the linear adaptors for each modality. $\mathbf{e}_*$ ($* \in \{m, v, i\}$) denotes the embedding of music, video, and image produced by ImageBind, while $\mathbf{e}_\#$ ($\# \in \{t, q, w\}$) are the LLM's text embeddings of input, query, and output. 


However, the pooled single embedding from ImageBind may fail to capture the detailed information required to effectively interact with other modalities in downstream music understanding tasks, particularly for music and video modalities. This limitation arises because such multimodal encoders prioritize coarse-grained representations for cross-modal retrieval tasks~\cite{DBLP:journals/corr/abs-2211-01324}. Additionally, as NExT-GPT relies solely on modality-specific adaptors to map each modality into LLM's embedding space, it may not effectively model interactions between modalities. These interactions were never pre-trained, and the LLM itself only employs uni-directional attention~\cite{DBLP:journals/corr/abs-2408-11039}.

To address these challenges, we propose two modules for music LLMs: \textbf{multi-sampled ImageBind embeddings} and \textbf{pre-LLM fusion Transformer}, as shown in Fig.~\ref{fig:model}. The former leverages embeddings from multiple clips sampled by ImageBind without pooling, while the latter introduces a Transformer to globally integrate and align information across modalities before feeding it into the LLM. Formally, the proposed model is defined as
\begin{align}
    \mathcal{P}(w_{n}|\mathbb{X}_m, \mathbb{X}_v, \mathbb{X}_i, \mathbb{X}_t, \mathbb{Q}, \mathbb{W}_{1:n-1}) = \nonumber \\
    \mathcal{LLM}(\mathcal{T}( \mathcal{A}_m(\{\mathbf{e}_m\}_{1:\mathrm{N}_m}), \mathcal{A}_v(\{\mathbf{e}_v\}_{1:\mathrm{N}_v}), \nonumber \\ 
    \mathcal{A}_i(\{\mathbf{e}_i\}_{1:\mathrm{N}_i}), \{\mathbf{e}_t\}), \{\mathbf{e}_q\}, \{\mathbf{e}_w\}_{1:n-1}).
\end{align}
Here, $\mathcal{T}$ represents the pre-LLM fusion Transformer, and $\mathbf{e}_*$ ($* \in \{m, v, i\}$) from Eq.~\ref{eq:nextgpt} is reformulated as $\{\mathbf{e}_*\}_{1:\mathrm{N}_*}$ to incorporate multi-sampled ImageBind embeddings for each modality, where $\mathrm{N}_*$ denotes the number of sampled clips for a given modality. With these components, the multi-sampled ImageBind embeddings preserve finer-grained information for each modality, which is expected to enhance music understanding tasks (Sec.~\ref{sec:eval}). The pre-LLM fusion Transformer uses bidirectional attention to encode cross-modal dependencies, enhancing the interactions across music and other modalities, aiming to improve multimodal music understanding tasks (Sec.~\ref{sec:mmmu}).

\begin{table}[t!]
    \centering
    \resizebox{\linewidth}{!}{
    \begin{tabular}{lll}
    \toprule
        \textbf{Dataset} & \textbf{Used for} & \textbf{In$\xrightarrow{}$out modality} \\
    \toprule
        COCO & Train stage 1 & I+T$\xrightarrow{}$T \\
        Music4way & Train stage 2 & I+T$\xrightarrow{}$T \\
        Music4way & Train stages 1 \& 2 & V+T$\xrightarrow{}$T \\
        Music4way & Train stages 1 \& 2 & M+T$\xrightarrow{}$T \\
        Alpaca & Train stage 2 & T$\xrightarrow{}$T \\
        MusicQA & Train stage 2 & M+T$\xrightarrow{}$T \\
        MusicCaps & Train stage 2 & M+T$\xrightarrow{}$T \\
        Music4way-MI2T & Train stage 2 & M+I+T$\xrightarrow{}$T \\
        Music4way-MV2T & Train stage 2 & M+V+T$\xrightarrow{}$T \\
    \bottomrule
    \end{tabular}
    }
    \caption{Overview of training data. M: Music; I: Image; V: Video; T: Text.}
    \label{tab:data}
\end{table}

\begin{table*}[t]
    \centering
    \resizebox{\linewidth}{!}{
    \begin{tabular}{l|rrr|rrr|rrr}
    \toprule
    \multirow{2}{*}{\textbf{Model}} & \multicolumn{3}{c|}{\textbf{MusicQA}} & \multicolumn{3}{c|}{\textbf{MusicCaps}} & \multicolumn{3}{c}{\textbf{Music4way-MusicCaps}} \\
    & \textbf{B-1} & \textbf{R-L} & \textbf{BERTS} & \textbf{B-1} & \textbf{R-L} & \textbf{BERTS} & \textbf{B-1} & \textbf{R-L} & \textbf{BERTS} \\
    \toprule
    \textbf{SALMONN}~\cite{DBLP:conf/iclr/TangYSC000M024} & $^\dagger$28.7 & $^\dagger$35.4 & $^\dagger$90.3 & $^\dagger$19.7 & $^\dagger$19.1 & $^\dagger$86.9 & 19.1 & 20.0 & 87.0 \\
    \textbf{MU-LLaMA}~\cite{DBLP:conf/icassp/LiuHSS24} & $^\dagger$29.7 & $^\dagger$33.1 & $^\dagger$89.9 & $^{*\dagger}$9.6 & $^{*\dagger}$16.2 & $^{*\dagger}$86.8 & 15.1 & 27.6 & 88.3 \\
    \textbf{OpenMU}~\cite{DBLP:journals/corr/abs-2410-15573} & $^\dagger$24.5 & $^\dagger$25.5 & $^\dagger$88.6 & $^\dagger$23.9 & $^\dagger$19.4 & $^\dagger$86.6 & -- & -- & -- \\
    \textbf{MusiLingo sft. w/ MusicCaps}~\cite{deng-etal-2024-musilingo} & -- & -- & -- & -- & $^\dagger$\textbf{21.7} & $^\dagger$86.8 & -- & -- & -- \\
    \textbf{NExT-GPT}~\cite{DBLP:conf/icml/Wu0Q0C24} & 23.3 & 26.0 & 87.6 & 16.5 & 14.0 & 84.0 & 16.6 & 17.2 & 86.7 \\
    \textbf{M$^2$UGen}~\cite{DBLP:journals/corr/abs-2311-11255} & $^\dagger$29.1 & $^\dagger$37.9 & $^\dagger$90.5 & $^{*\dagger}$14.4 & $^{*\dagger}$16.4 & $^{*\dagger}$86.5 & $^{*\dagger}$13.1 & $^{*\dagger}$26.0 & $^{*\dagger}$87.6 \\
    \textbf{NExT-GPT w/ M$^2$UGen} & $^\dagger$34.0 & $^\dagger$39.6 & $^\dagger$91.2 & 12.4 & 16.1 & 86.9 & $^{*\dagger}$23.2 & $^{*\dagger}$36.7 & $^{*\dagger}$91.3 \\
    \textbf{NExT-GPT w/ Music4way} & $^\dagger$34.1 & $^\dagger$39.2 & $^\dagger$91.3 & $^\dagger$25.0 & $^\dagger$21.0 & $^\dagger$87.2 & $^\dagger$39.1 & $^\dagger$46.8 & $^\dagger$93.0 \\
    \rowcolor{gray!20}
    \textbf{DeepResonance-$\alpha$ (ours)} & $^\dagger$\textbf{35.1} & $^\dagger$\textbf{40.8} & $^\dagger$\textbf{91.6} & $^\dagger$\textbf{26.0} & $^\dagger$\textbf{21.6} & $^\dagger$\textbf{87.3} & $^\dagger$\textbf{40.9} & $^\dagger$\textbf{48.4} & $^\dagger$\textbf{93.3} \\
    \rowcolor{gray!20}
    \textbf{DeepResonance-$\beta$ (ours)} & $^\dagger$\textbf{35.6} & $^\dagger$\textbf{41.1} & $^\dagger$\textbf{91.6} & $^\dagger$\textbf{25.8} & $^\dagger$\textbf{21.6} & $^\dagger$\textbf{87.3} & $^\dagger$\textbf{39.9} & $^\dagger$\textbf{47.8} & $^\dagger$\textbf{93.2} \\
    \bottomrule
    \end{tabular}
    }
    \caption{\textbf{Results on MusicQA, MusicCaps, and Music4way-MusicCaps.} The top two performances are highlighted in \textbf{bold}. ``*'' denotes the test data was included in the corresponding model's training set. ``$^\dagger$'' indicates supervised settings. ``B-1'', ``R-L'', and ``BERTS'' denote BLEU-1, ROUGE-L, and BERTScore, respectively.}
    \label{tab:mu}
\end{table*}

\section{Experiments and Results}
\label{sec:eval}
Following the training strategy of NExT-GPT and M$^2$UGen, we train DeepResonance in two stages. Table~\ref{tab:data} summarizes all datasets used for training, with training details provided in Appx.~\ref{appx:td}. Subsequently, we evaluate DeepResonance across three music understanding tasks and three multimodal music understanding tasks in supervised settings. Additionally, we assess the model's zero-shot performance on out-of-domain datasets and conduct an ablation study to demonstrate the effectiveness of each proposed component.  Results are reported using BLEU~\cite{papineni-etal-2002-bleu},\footnote{Following prior studies, we primarily report BLEU-1, while BLEU-1 and BLEU are shown in Appx.~\ref{appx:mu},~\ref{appx:mmu},~\ref{appx:zs}, and~\ref{appx:as}.} ROUGE-L~\cite{lin-2004-rouge}, and BERTScore~\cite{DBLP:conf/iclr/ZhangKWWA20}.



\subsection{Baselines and Ours}
\label{sec:baseline}
Below are the baselines and ours that we compare.

\noindent Existing baseline models: \textbf{SALMONN}~\cite{DBLP:conf/iclr/TangYSC000M024}, \textbf{MU-LLaMA}~\cite{DBLP:conf/icassp/LiuHSS24}, \textbf{NExT-GPT}~\cite{DBLP:conf/icml/Wu0Q0C24}, \textbf{M$^2$UGen}~\cite{DBLP:journals/corr/abs-2311-11255}, \textbf{MusiLingo}~\cite{deng-etal-2024-musilingo} and \textbf{OpenMU}~\cite{DBLP:journals/corr/abs-2410-15573} (details in Appx.~\ref{appx:baseline}).

\noindent \textbf{NExT-GPT w/ M$^2$UGen}: We train a NExT-GPT model on the same data as M$^2$UGen, excluding MusicCaps, as its test split was inadvertently included in M$^2$UGen's training data.


\noindent \textbf{NExT-GPT w/ Music4way}: We train a NExT-GPT model using the Music4way datasets (see Sec.~\ref{sec:4way}), a more extensive and 4-way aligned dataset compared to M$^2$UGen's training data. We exclude Music4way-MI2T and Music4way-MV2T (Table~\ref{tab:data}), aiming to evaluate the benefits of them.

\noindent \textbf{DeepResonance (Ours)}: The models introduced in Sec.~\ref{sec:model}, built upon NExT-GPT and trained with datasets as shown in Table~\ref{tab:data}. We introduce two DeepResonance variants: DeepResonance-$\alpha$, trained without the pre-LLM fusion Transformer, and DeepResonance-$\beta$, which includes it. The following sections examine their effectiveness, as the pre-LLM fusion Transformer is specifically designed for multimodal music understanding tasks (Sec.~\ref{sec:mmmu}). We also present the performance of DeepResonance models trained using the Mistral- and LLaMA-3-based NExT-GPT in Appx.~\ref{appx:mu}.


\begin{table*}[t]
    \centering
    \resizebox{0.785\linewidth}{!}{
    \begin{tabular}{l|rrr|rrr|rrr}
    \toprule
    \multirow{2}{*}{\textbf{Model}} & \multicolumn{3}{c|}{\textbf{Music4way-MI2T}} & \multicolumn{3}{c|}{\textbf{Music4way-MV2T}} & \multicolumn{3}{c}{\textbf{Music4way-Any2T}} \\
    & \textbf{B-1} & \textbf{R-L} & \textbf{BERTS} & \textbf{B-1} & \textbf{R-L} & \textbf{BERTS} & \textbf{B-1} & \textbf{R-L} & \textbf{BERTS} \\
    \toprule
    \textbf{NExT-GPT}~\cite{DBLP:conf/icml/Wu0Q0C24} & 26.7 & 21.3 & 85.2 & 26.5 & 21.0 & 84.8 & 25.4 & 23.4 & 86.6 \\
    \textbf{M$^2$UGen}~\cite{DBLP:journals/corr/abs-2311-11255} & $^*$31.7 & $^*$26.4 & $^*$87.1 & $^*$31.7 & $^*$25.9 & $^*$86.8 & $^*$20.8 & $^*$21.5 & $^*$87.3 \\
    \textbf{NExT-GPT w/ M$^2$UGen} & $^*$33.8 & $^*$27.3 & $^*$88.1 & $^*$34.6 & $^*$27.3 & $^*$88.1 & $^*$26.3 & $^*$28.5 & $^*$89.4 \\
    \textbf{NExT-GPT w/ Music4way} & 24.2 & 22.0 & 85.4 & 25.0 & 22.5 & 85.5 & 29.4 & 27.0 & 88.4 \\
    \rowcolor{gray!20}
    \textbf{DeepResonance-$\alpha$ (ours)} & \textbf{48.7} & \textbf{36.2} & \textbf{90.2} & \textbf{48.9} & \textbf{36.5} & \textbf{90.2} & \textbf{37.2} & \textbf{29.6} & \textbf{89.5} \\
    \rowcolor{gray!20}
    \textbf{DeepResonance-$\beta$ (ours)} & \textbf{49.2} & \textbf{36.8} & \textbf{90.2} & \textbf{49.0} & \textbf{36.8} & \textbf{90.3} & \textbf{33.5} & \textbf{27.4} & \textbf{88.7} \\
    \bottomrule
    \end{tabular}
    }
    \caption{\textbf{Results on Music4way-MI2T, Music4way-MV2T, and Music4way-Any2T.} The top two performances are highlighted in \textbf{bold}. ``*'' denotes the test data was included in the corresponding model's training set.}
    \label{tab:mmu}
\end{table*}

\begin{figure*}[t]
    \centering
    \begin{subfigure}{0.3\linewidth}
        \centering
        \includegraphics[width=\linewidth]{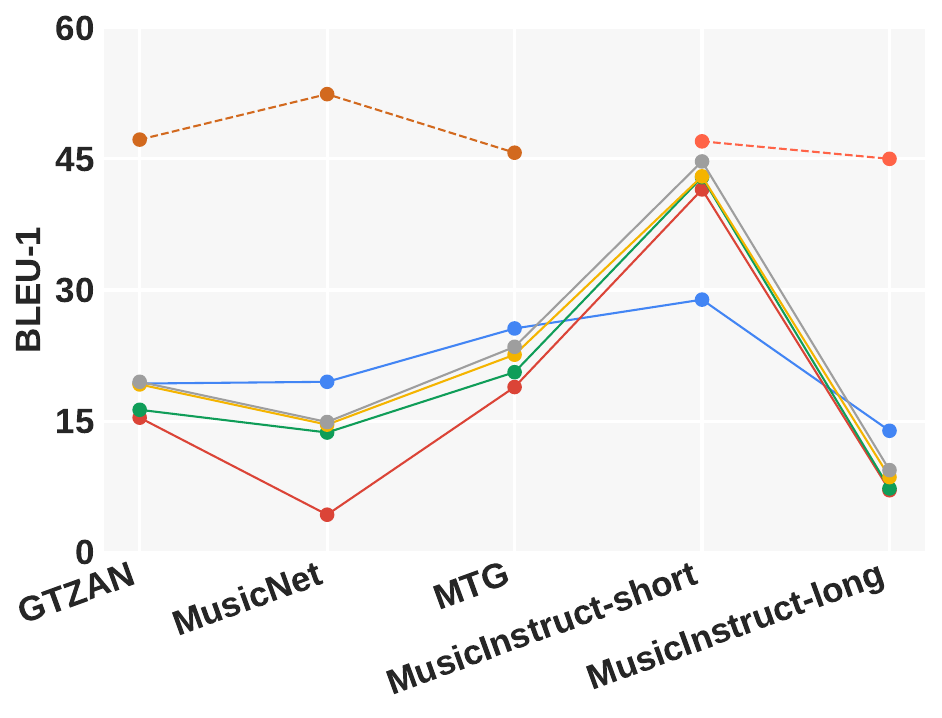}
    \end{subfigure}
    \hfill
    \begin{subfigure}{0.3\linewidth}
        \centering
        \includegraphics[width=\linewidth]{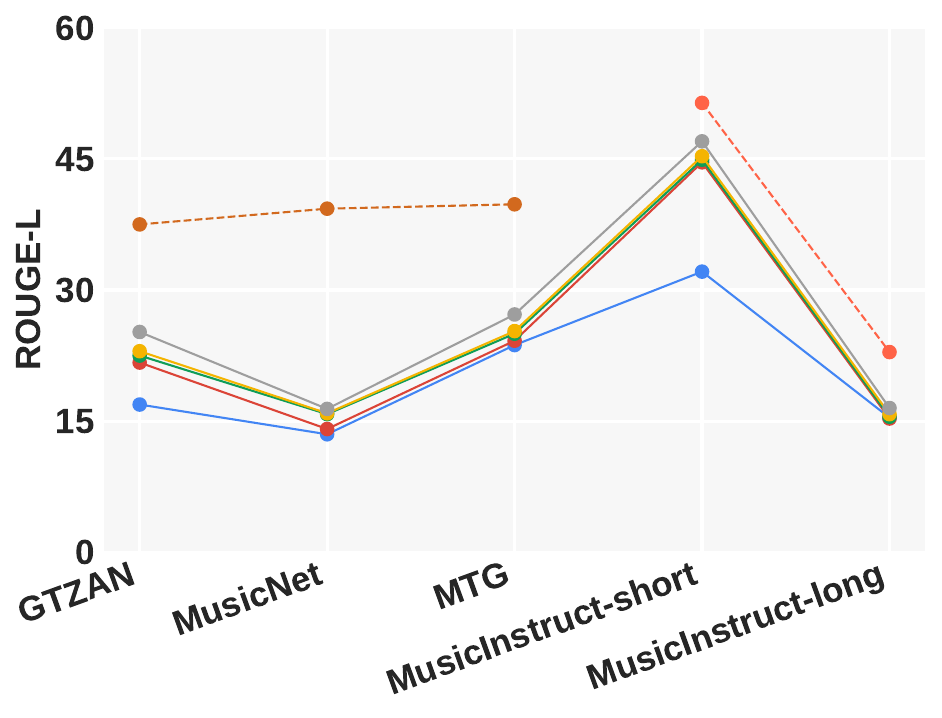}
    \end{subfigure}
    \hfill
    \begin{subfigure}{0.3\linewidth}
        \centering
        \includegraphics[width=\linewidth]{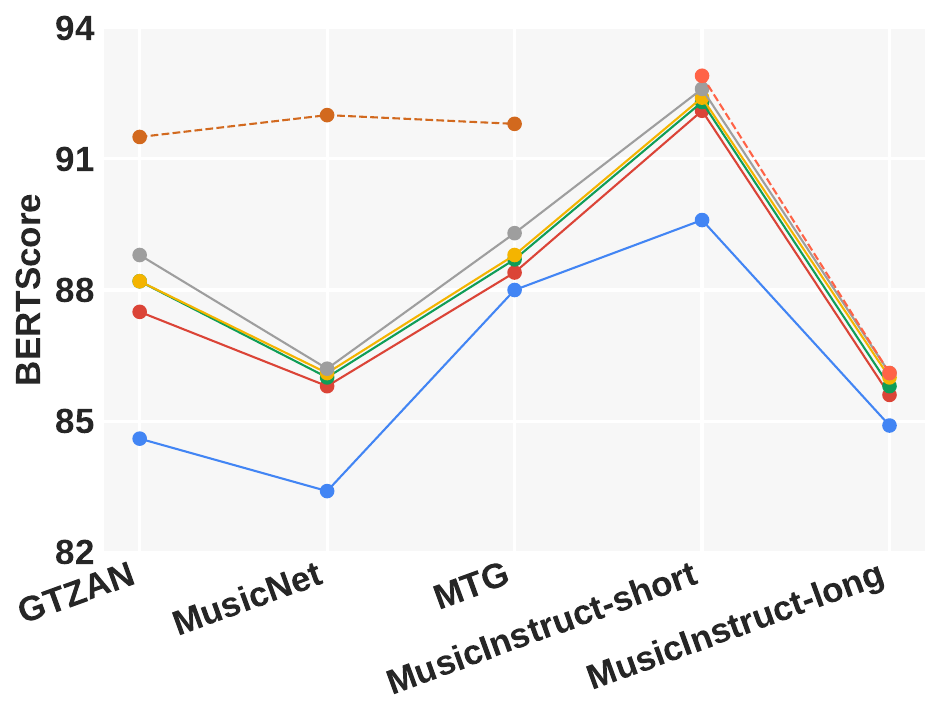}
    \end{subfigure}
    \hfill
    \begin{subfigure}{\linewidth}
        \centering
        \includegraphics[width=0.6\linewidth]{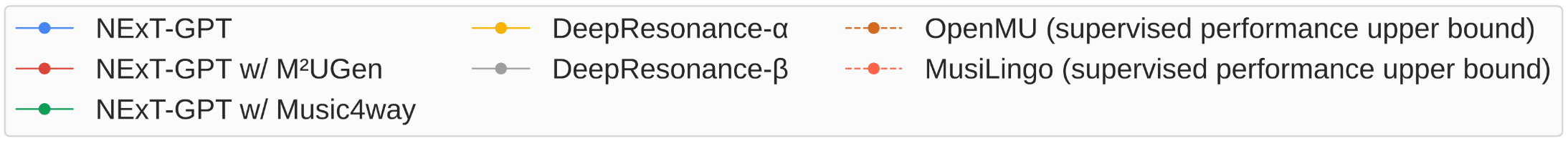}
    \end{subfigure}
    \caption{Zero-shot evaluation on GTZAN, MusicNet, MTG-Jamendo, MusicInstruct-short, and MusicInstruct-long.}
    \label{fig:zs}
\end{figure*}

\subsection{Music Understanding Tasks}
Table~\ref{tab:mu} reports the performance of all models introduced in Sec.\ref{sec:baseline} on music understanding tasks (music + text $\xrightarrow{}$ text) using MusicQA\cite{DBLP:conf/icassp/LiuHSS24}, MusicCaps~\cite{DBLP:journals/corr/abs-2301-11325}, and our constructed Music4way-MusicCaps\footnote{A music captioning subset of the Music4way test split.}.

First, we observe that DeepResonance outperforms baseline models on three datasets, demonstrating the effectiveness of our proposed training datasets and model architecture. Second, we find that applying the same training data to the NExT-GPT framework yields a better performance than using the M$^2$UGen framework, suggesting that NExT-GPT is a more suitable backbone model. Third, by expanding M$^2$UGen's training data with our constructed Music4way dataset, we observe an improvement in performance, further validating the effectiveness of the Music4way dataset. Finally, the comparison between DeepResonance-$\alpha$ and DeepResonance-$\beta$ shows a comparable performance, indicating that the pre-LLM fusion Transformer is not crucial for music + text $\xrightarrow{}$ text tasks. This aligns with expectations, as the pre-LLM fusion Transformer is designed for multimodal music understanding, whereas these tasks involve only a single modality (music) alongside the text query.


\begin{figure*}[t]
    \centering
    \includegraphics[width=0.95\linewidth]{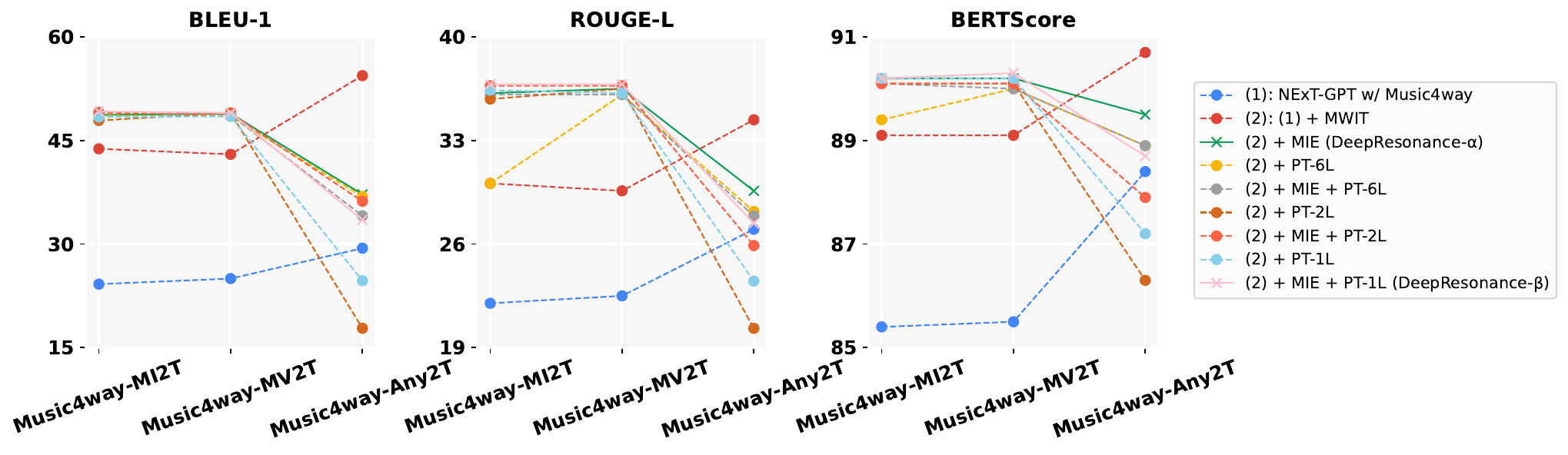}
    \caption{Ablation study on Music4way-MI2T, Music4way-MV2T, and Music4way-Any2T. ``PT-*L'' indicates the number of layers used in the pre-LLM fusion Transformer. See Appx.~\ref{appx:as} for result details.}
    \label{fig:as1}
\end{figure*}

\begin{figure}[t]
    \centering
    \includegraphics[width=0.95\linewidth]{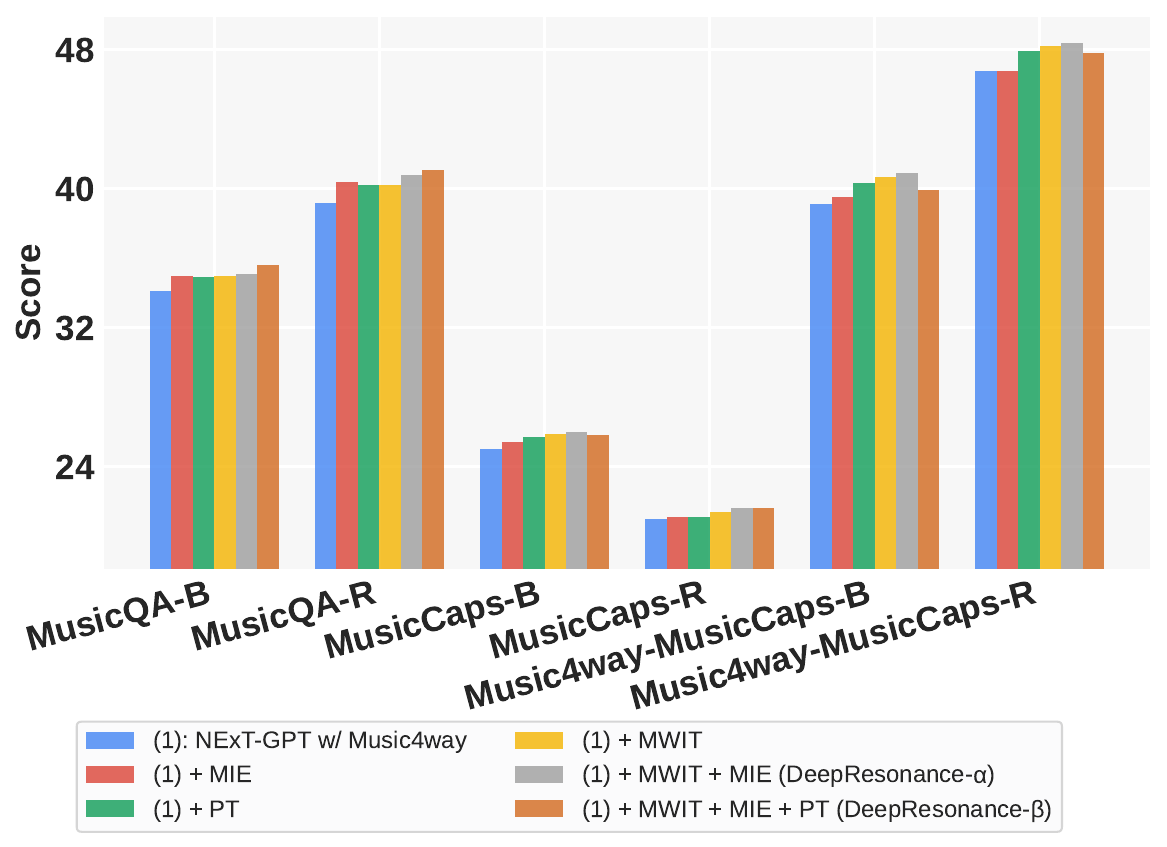}
    \caption{Ablation study on MusicQA, MusicCaps, and Music4way-MusicCaps. See Appx.~\ref{appx:as} for result details.}
    \label{fig:as2}
\end{figure}

\subsection{Multimodal Music Understanding Tasks}
\label{sec:mmmu}
Table~\ref{tab:mmu} lists the results on our proposed multimodal music understanding tasks (music + image/video + text $\xrightarrow{}$ text), including Music4way-MI2T, Music4way-MV2T, and Music4way-Any2T. We compare the performance of DeepResonance with baseline models capable of processing multiple modalities—music, text, image, and video—such as the M$^2$UGen and NExT-GPT-based models.

First, we observe that Music4way-MI2T and Music4way-MV2T represent novel downstream tasks for which no existing baseline models are inherently equipped. Through supervised fine-tuning with our curated training data for these datasets, the DeepResonance models gain the ability to generate unified multimodal captions successfully. Second, on Music4way-Any2T, which features flexible inputs and open-ended question-answer pairs, the baseline models perform poorly. Their generalization remains weaker than DeepResonance models, highlighting their limitations in handling diverse input patterns. Third, comparing DeepResonance-$\alpha$ and DeepResonance-$\beta$, we find that the latter demonstrates a superior performance on Music4way-MI2T and Music4way-MV2T. This indicates that the pre-LLM fusion Transformer, with its additional parameters, effectively integrates multiple modalities, thereby improving the supervised performance in structured multimodal music understanding tasks. However, DeepResonance-$\beta$ exhibits reduced robustness on Music4way-Any2T, which highlights a trade-off, as the model's increased complexity may hinder its adaptability with limited instruction tuning data.


\subsection{Zero-shot Evaluation}
Fig.~\ref{fig:zs} present the zero-shot performance of DeepResonance and baselines on music understanding benchmarks, including GTZAN~\cite{DBLP:conf/ismir/Tzanetakis01}, MusicNet~\cite{DBLP:conf/iclr/ThickstunHK17}, MTG-Jamendo~\cite{bogdanov2019mtg}, MusicInstruct-short, and MusicInstruct-long~\cite{deng-etal-2024-musilingo}. We adopt the benchmark settings outlined in OpenMU-bench~\cite{DBLP:journals/corr/abs-2410-15573}, combining captioning and reasoning test sets where separate splits exist. 
All NExT-GPT-based models without exposure to test data during training are compared, to ensure the zero-shot configurations. Detailed results are provided in the Appx.~\ref{appx:zs}.

First, DeepResonance-$\beta$ achieves the best performance across all five benchmarks in terms of ROUGE-L and BERTScore, demonstrating the effectiveness of the proposed training data and model architecture. Second, DeepResonance-$\beta$ outperforms DeepResonance-$\alpha$, highlighting the pre-LLM fusion Transformer's effectiveness in improving inference on unseen data. Referring back to Sec.~\ref{sec:mmmu}, we recommend DeepResonance-$\alpha$ for tasks with flexible inputs and DeepResonance-$\beta$ for other scenarios. Finally, DeepResonance's zero-shot performance approaches the supervised upper bounds of MusiLingo on MusicInstruct in terms of BERTScore, demonstrating the out-of-domain generalization capabilities of DeepResonance.


\subsection{Ablation Study}
\label{sec:ablation}
Figs.~\ref{fig:as1} and~\ref{fig:as2} present the results of ablation studies evaluating the effectiveness of key components in the proposed methods, including Music4way-MI2T and Music4way-MV2T instruction tuning data (\textbf{MWIT}), multi-sampled ImageBind embeddings (\textbf{MIE}), and the pre-LLM fusion Transformer (\textbf{PT}). For music + text $\xrightarrow{}$ text tasks (Fig.~\ref{fig:as2}), we observe that MWIT, MIE, and PT each contribute positively to performance. However, when combined, PT does not consistently complement the other two components across all three benchmarks, with MWIT + MIE (DeepResonance-$\alpha$) yielding the consistent improvements. For multimodal music understanding tasks (music + image/video + text → text), we compare settings with and without MIE and PT in Fig.~\ref{fig:as1}, as MWIT serves as the in-domain data for these tasks. Integrating MIE and PT enhances performance, with PT proving most effective when limited to a single Transformer layer. This highlights the effectiveness of MIE and PT while suggesting that increasing PT’s parameters may lead to overfitting on limited instruction tuning data. Moreover, PT negatively impacts performance on Music4way-Any2T (refer to Sec.~\ref{sec:mmmu}).

\begin{table}[t]
    \centering
    \resizebox{\linewidth}{!}{
    \begin{tabular}{l|rrr|rrr|rrr}
    \toprule
    \multirow{2}{*}{\textbf{Model}} & \multicolumn{3}{c|}{\textbf{MusicQA}} & \multicolumn{3}{c|}{\textbf{MusicCaps}} & \multicolumn{3}{c}{\textbf{Music4way-MusicCaps}} \\
    & \textbf{B-1} & \textbf{R-L} & \textbf{BERTS} & \textbf{B-1} & \textbf{R-L} & \textbf{BERTS} & \textbf{B-1} & \textbf{R-L} & \textbf{BERTS} \\
    \toprule
    DR-$\alpha$ & 35.1 & 40.8 & 91.6 & 26.0 & 21.6 & 87.3 & 40.9 & 48.4 & 93.3 \\
     w/o v.c. & 35.0 & 40.4 & 91.4 & 25.5 & 21.2 & 87.2 & 40.7 & 48.1 & 93.2 \\
     w/o m.f. & 34.9 & 40.3 & 91.5 & 25.9 & 21.3 & 87.3 & 40.5 & 48.0 & 93.2 \\
    \hline
    DR-$\beta$ & 35.6 & 41.1 & 91.6 & 25.8 & 21.6 & 87.3 & 39.9 & 47.8 & 93.2 \\
     w/o v.c. & 35.2 & 40.7 & 91.5 & 25.8 & 21.2 & 87.2 & 39.5 & 46.7 & 93.0 \\
     w/o m.f. & 35.2 & 40.2 & 91.4 & 25.8 & 21.1 & 87.3 & 40.0 & 48.0 & 93.2 \\
    \bottomrule
    \end{tabular}
    }
    \caption{Ablation study on the impact of decoder-side visual and musical contents. See details in Appx.~\ref{appx:as}. ``v.c.'': visual captions; ``m.f.'': low-level musical features.}
    \label{tab:as}
\end{table}

We also conduct an ablation study on the multi-way unified captions of Music4way-MI2T and Music4way-MV2T. Referring to Fig.\ref{fig:data}, we remove image or video captions to isolate the contribution of visual contents, and similarly remove low-level musical features to assess the impact of detailed musical information. As shown in Table~\ref{tab:as}, removing either leads to a slight performance drop, suggesting that the unified target formats of both contents enhance music understanding capability. However, comparisons in Fig.\ref{fig:as2} indicate that using MWIT and applying our proposed MIE or PT methods have a greater impact than the specific target-side content of the captions.


\section{Conclusion}
In this study, we introduced DeepResonance, a multimodal music understanding LLM capable of comprehending music through its connections with other modalities, such as image and video. To train and evaluate DeepResonance, we developed new music-centric multi-way instruction-following datasets. In addition, we proposed modules designed to enhance music-centric multimodal instruction tuning. Empirical results highlight the effectiveness of DeepResonance across six music understanding tasks and zero-shot scenarios. Future work will explore more refined instruction tuning datasets to improve the model’s generalization capabilities for music understanding tasks.

\section*{Limitations}
The proposed methods have the following limitations:
\begin{itemize}
    \item[(1)] The input music training data is mostly limited to clips shorter than 30 seconds, with a significant portion (e.g., AudioSet) being 10 seconds. This may restrict the fine-tuned models' performance on longer music sequences. Additionally, the dataset’s image frames are directly extracted from videos, assuming relevance within short clips (10s). For longer videos, selecting the most representative frames—such as cover images—should be explored in future work, once more licensed long-form music and video clips become available. (This is beyond the scope of this paper, as scene transitions within the 10-second AudioSet clips are rare; therefore, randomly selecting a frame is sufficient.)
    \item[(2)] While the proposed methods perform well on supervised datasets included in the training data, further the enhancing generalization capability to out-of-domain (distribution-shifted) music remains an open challenge. 
    \item[(3)] Generating instruction tuning data with LLMs is a well-established and widely accepted approach, as seen in Self-Instruct~\cite{wang-etal-2023-self-instruct}. Our instruction tuning data construction process is relatively simple, ensuring the overall reliability. However, as with any LLM-generated data, biases may exist, and users should be mindful of potential biases.
\end{itemize}


\section*{Ethical Considerations}
In this study, we leveraged publicly available datasets (without licensing issues) to create new datasets for multimodal music understanding. The newly generated content consists solely of text produced by LLMs such as GPT-4o mini, with no originally generated music, images, or videos. We fine-tuned the multimodal music understanding model through instruction tuning. While the model has been adapted to a specific domain, it may still generate hallucinations or biased content due to the nature of LLM-based text generation. Users should exercise caution when using the generated content, be aware of the potential risks associated with LLM outputs, and implement content safety checks as a post-processing measure.



\bibliography{anthology,custom}


\appendix

\begin{table*}[t]
    \centering
    \resizebox{\linewidth}{!}{
    \begin{tabular}{lrrlll}
    \toprule
        \textbf{Dataset} & \textbf{\#Instance\_train} & \textbf{\#Instance\_test} & \textbf{Used for} & \textbf{In$\xrightarrow{}$out modality} & \textbf{Task} \\
    \toprule
        COCO & 82,783 & -- & Train stage 1 & I+T$\xrightarrow{}$T & Image captioning \\
        Music4way & 59,128 & -- & Train stage 2 & I+T$\xrightarrow{}$T & Image captioning \\
        Music4way & 59,128 & -- & Train stages 1 \& 2 & V+T$\xrightarrow{}$T & Video captioning \\
        Music4way & \multirow{2}{*}{59,128} & \multirow{2}{*}{3,000} & \multirow{2}{*}{Train stages 1 \& 2, test} & \multirow{2}{*}{M+T$\xrightarrow{}$T} & \multirow{2}{*}{Music captioning} \\
        (Music4way-MusicCaps) \\
        Alpaca & 51,974 & -- & Train stage 2 & T$\xrightarrow{}$T & Text question-answering \\
        MusicQA & 70,011 & 5,040 & Train stage 2, test & M+T$\xrightarrow{}$T & Music captioning and question-answering \\
        MusicCaps & 2,640 & 2,839 & Train stage 2, test & M+T$\xrightarrow{}$T & Music captioning \\
        Music4way-MI2T & 59,128 & 3,000 & Train stage 2, test & M+I+T$\xrightarrow{}$T & Multimodal captioning \\
        Music4way-MV2T & 59,128 & 3,000 & Train stage 2, test & M+V+T$\xrightarrow{}$T & Multimodal captioning \\
        Music4way-Any2T & 59,128 & 3,000 & Test & M+I/V+T$\xrightarrow{}$T & Multimodal question-answering \\
        GTZAN & -- & 1,406 & Test & M+T$\xrightarrow{}$T & Music captioning and question-answering \\
        MusicNet & -- & 140 & Test & M+T$\xrightarrow{}$T & Music captioning \\
        MusicInstruct-long & -- & 16,658 & Test & M+T$\xrightarrow{}$T & Music captioning \\
        MusicInstruct-short & -- & 13,935 & Test & M+T$\xrightarrow{}$T & Music captioning \\
        MTG & -- & 25,452 & Test & M+T$\xrightarrow{}$T & Music captioning and question-answering \\
    \bottomrule
    \end{tabular}
    }
    \caption{Overview of datasets used for training and evaluation. M: music; I: image; V: video; T: text.}
    \label{tab:data2}
\end{table*}

\begin{table*}[t!]
    \centering
    \resizebox{\linewidth}{!}{
    \begin{tabular}{lrrrrrrrr}
    \toprule
    \textbf{Model} & \textbf{BLEU-1} & \textbf{BLEU} & \textbf{ROUGE-P} & \textbf{ROUGE-R} & \textbf{ROUGE-F1} & \textbf{BERT-P} & \textbf{BERT-R} & \textbf{BERT-F1} \\
    \toprule
    \textit{MusicQA} \\
    \textbf{DeepResonance-$\alpha$} & 35.1 & 15.1 & 51.0 & 38.6 & 40.8 & 92.5 & 90.7 & 91.6 \\
    \ \ \ \ \ \ \ \ w/o music & 36.8 & 15.5 & 47.1 & 41.4 & 40.7 & 91.7 & 90.8 & 91.2 \\
    \textbf{DeepResonance-$\beta$} & 35.6 & 15.3 & 51.3 & 39.0 & 41.1 & 92.5 & 90.8 & 91.6 \\
    \ \ \ \ \ \ \ \ w/o music & 38.5 & 15.5 & 44.2 & 42.6 & 40.5 & 91.3 & 90.9 & 91.1 \\
    \hline
    \textit{MusicCaps} \\
    \textbf{DeepResonance-$\alpha$} & 26.0 & 3.0 & 23.4 & 21.8 & 21.6 & 87.6 & 87.1 & 87.3 \\
    \ \ \ \ \ \ \ \ w/o music & 22.6 & 1.0 &  21.7 & 19.1 & 19.0 & 87.6 & 86.1 & 86.8 \\
    \textbf{DeepResonance-$\beta$} & 25.8 & 2.8 & 23.6 & 21.4 & 21.6 & 87.5 & 87.1 & 87.3 \\
    \ \ \ \ \ \ \ \ w/o music & 23.3 & 1.9 &  20.3 & 20.4 & 19.3 & 86.6 & 86.2 & 86.4 \\
    \hline
    \textit{Music4way-MusicCaps} \\
    \textbf{DeepResonance-$\alpha$} & 40.9 & 19.9 & 57.8 & 47.5 & 48.4 & 94.0 & 92.6 & 93.3 \\
    \ \ \ \ \ \ \ \ w/o music & 35.7 & 13.0 & 45.2 & 43.0 & 41.6 & 90.5 & 92.0 & 91.2 \\
    \textbf{DeepResonance-$\beta$} & 39.9 & 19.3 & 57.3 & 47.3 & 47.8 & 93.9 & 92.5 & 93.2 \\
    \ \ \ \ \ \ \ \ w/o music & 33.4 & 14.1 & 57.8 & 40.4 & 44.9 & 93.6 & 91.0 & 92.3 \\
    \hline
    \textit{Music4way-MI2T} \\
    \textbf{DeepResonance-$\alpha$} & 48.7 & 16.4 & 37.1 & 36.8 & 36.2 & 90.3 & 90.2 & 90.2 \\
    \ \ \ \ \ \ \ \ w/o music \& image & 48.1 & 16.0 & 35.0 & 37.3 & 35.6 & 89.9 & 90.1 & 90.0 \\
    \textbf{DeepResonance-$\beta$} & 49.2 & 17.2 & 36.8 & 38.3 & 36.8 & 90.1 & 90.3 & 90.2 \\
    \ \ \ \ \ \ \ \ w/o music \& image & 42.1 & 13.2 & 30.1 & 38.4 & 32.3 & 88.1 & 88.9 & 88.5 \\
    \hline
    \textit{Music4way-MV2T} \\
    \textbf{DeepResonance-$\alpha$} & 48.9 & 16.6 & 37.4 & 37.0 & 36.5 & 90.3 & 90.2 & 90.2 \\
    \ \ \ \ \ \ \ \ w/o music \& video & 46.1 & 14.7 & 35.9 & 35.3 & 34.9 & 89.8 & 89.7 & 89.8 \\
    \textbf{DeepResonance-$\beta$} & 49.0 & 17.2 & 36.7 & 38.4 & 36.8 & 90.3 & 90.3 & 90.3 \\
    \ \ \ \ \ \ \ \ w/o music \& video & 42.0 & 13.0 & 30.1 & 38.4 & 32.3 & 88.1 & 89.0 & 88.5 \\
    \hline
    \textit{Music4way-Any2T} \\
    \textbf{DeepResonance-$\alpha$} & 37.2 & 6.0 & 36.0 & 25.7 & 29.6 & 90.8 & 88.3 & 89.5 \\
    \ \ \ \ \ \ \ \ w/o music, image \& video & 36.2 & 4.5 &  33.7 & 25.0 & 28.3 & 90.2 & 87.6 & 88.9 \\
    \textbf{DeepResonance-$\beta$} & 33.5 & 3.9 & 34.0 & 23.9 & 27.4 & 90.0 & 87.5 & 88.7 \\
    \ \ \ \ \ \ \ \ w/o music, image \& video & 33.0 & 3.7 & 33.5 & 23.6 & 27.2 & 89.9 & 87.3 & 88.5 \\
    \bottomrule
    \end{tabular}
    }
    \caption{Sanity check during the inference phase for music understanding and multimodal music understanding tasks.}
    \label{tab:sanity}
\end{table*}

\begin{table*}[t]
    \centering
    \resizebox{\linewidth}{!}{
    \begin{tabular}{l|p{\linewidth}}
    \toprule
        \multirow{7}{*}{Input} & \texttt{- Video Caption: A man is playing a song on a guitar and he is sitting on a table.} \\
        & \texttt{- Image Caption: The image shows a man playing a guitar.} \\
        & \texttt{- Music Caption: The music is a slow, heavy metal song with a dark and intense mood. The instruments used are electric guitar, bass guitar, drums, and vocals ...} \\
        & \texttt{- Music Features:} \\
        & \texttt{------ Tempo: [[105.26315789473684, 0.4619219529856875], ... ]} \\
        & \texttt{------ Chords: [[0.0, 1.5, 'C:maj'], [1.5, 3.7, 'F:maj'], ... ]} \\
        & \texttt{------ Downbeats: [[0.25, 3.0], [0.81, 4.0], [1.39, 1.0], [1.99, 2.0], ... ]} \\
        & \texttt{------ Key: [[9.83237405307591e-05, 8.27968106023036e-05, ...]} \\
        \hline
        \multirow{8}{*}{Instruction} & \texttt{Given the above information of video captions, image captions, music captions, and music features, generate a unified description that combines the elements of both the video and the music, taking into account the mood, style, and emotions conveyed by the captions and music features. The description should be cohesive and provide a holistic view of the content, reflecting how the visual and auditory components complement each other. Focus on creating a narrative that integrates the rhythm, harmony, and tonality of the music with the visual elements and storyline of the video.} \\
        \hline
        \multirow{9}{*}{Output (Unified Captions)} & \texttt{The video and music describe a fast-paced and energetic performance by a group of musicians, with a positive and upbeat mood. The music features a driving rhythm, a strong beat, and a variety of instruments, including drums, bass, electric guitar, and keyboards. The chords used in the music are major, with a key of A minor. The tempo ranges from 60 to 120 beats per minute, with a strong sense of energy and movement conveyed by the upbeat tone of the music. The downbeats occur every four beats, with a strong emphasis on the first and third beats of each measure. Overall, the visual and auditory components complement each other to create a cohesive and engaging experience for the viewer.} \\
    \bottomrule
    \end{tabular}
    }
    \caption{Prompting MPT-7B to generate unified captions for Music4way-MI2T and Music4way-MV2T.}
    \label{tab:mimv2t}
\end{table*}

\section{Discussion on Multi-way Alignment}
The 4-way alignment introduced in this work, which connects music, text, image, and video, is facilitated by pairing music with video and video with image. Each modality is further linked to text through captioning or feature extraction. Therefore, the 4-way relationship is constructed from several 2-way mappings, with any pair among four modalities being closely correlated as they stem from a single original music-video pair. Future work may develop finer-grained multi-way alignment for music understanding tasks. We encourage further discussion and research on how to establish improved multi-way relationships across different modalities.

\section{Training and Evaluation Details}
\label{appx:td}
Following the training strategy of NExT-GPT and M$^2$UGen, we train DeepResonance in two stages. In the first stage, we fine-tune only the parameters of the linear adaptors and the proposed pre-LLM fusion Transformer. This stage focuses on captioning tasks for music, image, and video modalities, utilizing images from COCO~\cite{DBLP:conf/eccv/LinMBHPRDZ14} and music and video clips from the constructed Music4way dataset.\footnote{We use COCO instead of Music4way for the image captioning task in the first stage, as empirical results indicate that using the larger image-text dataset, COCO, yields better performance.} In the second stage, we fine-tune the linear adaptors and the pre-LLM fusion Transformer while simultaneously performing LoRA-based fine-tuning~\cite{DBLP:conf/iclr/HuSWALWWC22} on Vicuna. This stage incorporates instruction tuning tasks using datasets including Alpaca~\cite{alpaca}, MusicCaps~\cite{DBLP:journals/corr/abs-2301-11325}, and MusicQA~\cite{DBLP:conf/icassp/LiuHSS24}\footnote{We use the ``fine-tune'' split of the MusicQA dataset and exclude the ``train'' split to avoid overlap with the test split of MusicCaps.}, along with our constructed Music4way, Music4way-MI2T, and Music4way-MV2T datasets. A summary of all datasets used for training and evaluation is provided in Table~\ref{tab:data}. As depicted in Fig.~\ref{fig:model}, instructions are fed directly into the text-LLM, bypassing the adaptors and the pre-LLM fusion Transformer, as they do not require interaction with the input information.

We fine-tune for $5$ and $2$ epochs in the first and second stages, respectively, utilizing a learning rate of $1\mathrm{e}{-4}$ and a batch size of $16$. Training is conducted on $8$ NVIDIA A100 GPUs (40GB each). For LoRA, the rank and alpha are both set to $32$, following NExT-GPT. We train the pre-LLM fusion Transformer with various layer configurations (see Sec.~\ref{sec:ablation}) and find that a single Transformer layer achieves the best performance. Regarding the trainable parameters, the linear adaptors, LoRA, pre-LLM fusion Transformer, and LLaMA embedding layers contain $4$M, $33$M, $157$M, and $262$M parameters, respectively, comprising 5.6\% of the total model parameters. Regarding the training budget, stage 1 took $25.2$ hours, while stage 2 took $20.3$ hours for DeepResonance-$\alpha$. For DeepResonance-$\beta$, the training times were $27.0$ hours for stage 1 and $22.7$ hours for stage 2.

For evaluation, we report the mean results from three inference runs and include BLEU~\cite{papineni-etal-2002-bleu}, ROUGE-L~\cite{lin-2004-rouge}, and BERTScore~\cite{DBLP:conf/iclr/ZhangKWWA20}, following the setup of M$^2$UGen~\cite{DBLP:journals/corr/abs-2311-11255}. We report BLEU-1, BLEU, ROUGE-L precision, ROUGE-L recall, ROUGE-L F1, BERTScore precision, BERTScore recall, and BERTScore F1 details in Appx.~\ref{appx:mu},~\ref{appx:mmu},~\ref{appx:zs}, and~\ref{appx:as}.

\section{Existing Baseline Models}
\label{appx:baseline}
\noindent \textbf{SALMONN}~\cite{DBLP:conf/iclr/TangYSC000M024}: A robust baseline model for audio understanding. It leverages a variety of audio datasets for training, including AudioCaps~\cite{DBLP:conf/naacl/KimKLK19}, WaveCaps~\cite{DBLP:journals/taslp/MeiMLKKZPZW24}, MusicNet~\cite{DBLP:conf/iclr/ThickstunHK17}, etc.


\noindent \textbf{MU-LLaMA}~\cite{DBLP:conf/icassp/LiuHSS24}: The first LLM instruction-tuned for music understanding with the MusicCaps and MusicQA as fine-tuning data.

\noindent \textbf{NExT-GPT}~\cite{DBLP:conf/icml/Wu0Q0C24}: The first any-to-any multimodal LLM trained on multimodal fine-tuning data, serving as the backbone for ours.

\noindent \textbf{M$^2$UGen}~\cite{DBLP:journals/corr/abs-2311-11255}: The first any-to-any multimodal LLM tailored to the music domain, trained on newly curated data derived from the music split of AudioSet~\cite{DBLP:conf/icassp/GemmekeEFJLMPR17}.

We also include results from a task-specific music understanding model, \textbf{MusiLingo}~\cite{deng-etal-2024-musilingo}, for comparison. We also report the performance of \textbf{OpenMU}~\cite{DBLP:journals/corr/abs-2410-15573}, a benchmark model for multiple music understanding datasets for supervised comparison. In zero-shot evaluations, OpenMU and MusiLingo serve as upper bounds for supervised performance, as they were directly trained on those benchmarks.

\section{Sanity check during the Inference Phase}
We perform a sanity check during the inference phase to assess the contribution of multimodal inputs to overall performance. Specifically, we retain only the textual inputs while removing all other modalities—music, image, and video—and evaluate the model on all six supervised music understanding tasks. The results, presented in Table~\ref{tab:sanity}, show a clear performance drop when multimodal inputs are excluded. This suggests that the DeepResonance models effectively leverage semantic information from non-text modalities. Moreover, the observed performance degradation highlights the utility and validity of our constructed Music4Way evaluation datasets for testing multimodal music LLMs.

\section{Construction Details and Data Examples of Music4way-MI2T and Music4way-MV2T}
\label{appx:mimv2t}
Table~\ref{tab:mimv2t} presents the specific templates used to prompt the MPT-7B model~\cite{MosaicML2023Introducing} to generate unified captions based on music, video, and image captions, along with low-level music features. Table~\ref{tab:mimv2tdata} provides data examples from the Music4way-MI2T and Music4way-MV2T datasets we constructed for multi-way instruction tuning.

\section{Construction Details and Data Examples of Music4way-Any2T}
\label{appx:any2t}
Table~\ref{tab:any2t} displays the templates used to prompt GPT-4o mini\footnote{\url{https://platform.openai.com/docs/models\#gpt-4o-mini}} to generate structured text comprising input, instruction, and output, based on music, video, and image captions, unified captions, and music features. Table~\ref{tab:any2tdata} presents data examples from the Music4way-Any2T dataset we constructed to evaluate the robustness and generalization capabilities of music LLMs.

\section{Detailed Results on Music Understanding Tasks}
\label{appx:mu}
Tables~\ref{tab:musicqa},~\ref{tab:musiccaps}, and~\ref{tab:music4way-musiccaps} report the detailed performance on all metrics of all models introduced in Sec.\ref{sec:baseline} on music understanding tasks (music + text $\xrightarrow{}$ text) using MusicQA\cite{DBLP:conf/icassp/LiuHSS24}, MusicCaps~\cite{DBLP:journals/corr/abs-2301-11325}, and our constructed Music4way-MusicCaps.

We also report the results of DeepResonance models based on LLaMA-3\footnote{We used LLaMA-3-8B-Instruct.}~\cite{DBLP:journals/corr/abs-2407-21783} and Mistral\footnote{We used Ministral-8B-Instruct-2410.}~\cite{DBLP:journals/corr/abs-2310-06825} in Tables~\ref{tab:musicqa},\ref{tab:musiccaps}, and\ref{tab:music4way-musiccaps}. Our default DeepResonance models built on Vicuna LLMs achieve the best overall performance, likely due to their smaller embedding layers, which are easier to fine-tune and adapt to multimodal contexts. Nevertheless, the Mistral- and LLaMA-3-based DeepResonance models still outperform their respective NExT-GPT counterparts trained on the Music4way dataset.

\section{Detailed Results on Multimodal Music Understanding Tasks}
\label{appx:mmu}
Tables~\ref{tab:music4way-mi2t},~\ref{tab:music4way-mv2t}, and~\ref{tab:music4way-any2t} list the result details on all metrics on our newly proposed multimodal music understanding tasks (music + image/video + text $\xrightarrow{}$ text), including Music4way-MI2T, Music4way-MV2T, and Music4way-Any2T.

\section{Detailed Results of Zero-shot Evaluation}
\label{appx:zs}
The detailed results of the zero-shot evaluation are presented in Tables~\ref{tab:gtzan},~\ref{tab:musicnet},~\ref{tab:musicinstructlong},~\ref{tab:musicinstructshort}, and~\ref{tab:mtg}.

\section{Detailed Results of Ablation Study}
\label{appx:as}
The detailed results of the ablation study are presented in Tables~\ref{tab:ablation1},~\ref{tab:ablation2},~\ref{tab:ablation3},~\ref{tab:ablation4},~\ref{tab:ablation5} and~\ref{tab:ablation6}.

\begin{table*}[t]
    \centering
    \resizebox{\linewidth}{!}{
    \begin{tabular}{l|p{\linewidth}}
    \toprule
        Input (Music4way-MI2T) & \texttt{<Music> <Image>} \\
        Input (Music4way-MV2T) & \texttt{<Music> <Video>} \\
        \hline
        \multirow{4}{*}{Instruction} & \texttt{Analyze the music by considering both its auditory and visual components. Describe the music in detail, incorporating its tempo, chords, downbeats, and key, while also reflecting on how these musical features align with the video or a key image from the video.} \\
        \hline
        \multirow{9}{*}{Output (Unified Captions)} & \texttt{The video and music describe a fast-paced and energetic performance by a group of musicians, with a positive and upbeat mood. The music features a driving rhythm, a strong beat, and a variety of instruments, including drums, bass, electric guitar, and keyboards. The chords used in the music are major, with a key of A minor. The tempo ranges from 60 to 120 beats per minute, with a strong sense of energy and movement conveyed by the upbeat tone of the music. The downbeats occur every four beats, with a strong emphasis on the first and third beats of each measure. Overall, the visual and auditory components complement each other to create a cohesive and engaging experience for the viewer.} \\
    \bottomrule
    \end{tabular}
    }
    \caption{Data examples of Music4way-MI2T and Music4way-MV2T for instruction tuning. ``<Music>'', ``<Image>'', and ``<Video>'' represent the paths to the respective files.}
    \label{tab:mimv2tdata}
\end{table*}

\begin{table*}[t]
    \centering
    \resizebox{\linewidth}{!}{
    \begin{tabular}{l|p{\linewidth}}
    \toprule
        \multirow{15}{*}{Input} & \texttt{Generate the output for the following input; the style of the input, instruction, and output may vary.} \\
        \\
        & \texttt{- Video Caption: A man is playing a song on a guitar and he is sitting on a table.} \\
        & \texttt{- Image Caption: The image shows a man playing a guitar.} \\
        & \texttt{- Music Caption: The music is a slow, heavy metal song with a dark and intense mood. The instruments used are electric guitar, bass guitar, drums, and vocals ...} \\
        & \texttt{- Music Features:} \\
        & \texttt{------ Tempo: [[105.26315789473684, 0.4619219529856875], ... ]} \\
        & \texttt{------ Chords: [[0.0, 1.5, 'C:maj'], [1.5, 3.7, 'F:maj'], ... ]} \\
        & \texttt{------ Downbeats: [[0.25, 3.0], [0.81, 4.0], [1.39, 1.0], [1.99, 2.0], ... ]} \\
        & \texttt{------ Key: [[9.83237405307591e-05, 8.27968106023036e-05, ...]} \\
        & \texttt{- Unified Caption: The video and music describe a fast-paced and energetic performance by a group of musicians, with a positive and upbeat mood. The music features a driving rhythm, a strong beat, and a variety of instruments, including drums, bass, ...} \\
        \hline
        \multirow{28}{*}{Instruction} & \texttt{Help generate input, instruction, and output triplets using the given paired music caption, video caption, image caption, music features, and unified caption. The unified caption includes all three captions as well as music features such as tempo, chords, downbeats, and key.} \\
        \\
        & \texttt{Guidelines:} \\
        & \texttt{1. Input: Should be a sentence that includes two or all three modalities: music (mandatory), video, and/or image. Music must be referred to as <Music>. Image must be referred to as <Image>. Video must be referred to as <Video>.} \\
        & \texttt{2. Instruction: Should be a text-based question or directive that requires generating a unified output based on the given inputs. It should guide the model to consider both the audio and the visual aspects, explaining how they interact to create a unified experience.} \\
        & \texttt{3. Output: Should be a textual response, potentially composed using the information from the music, video, image captions or the unified caption.} \\
        \\
        & \texttt{Example:} \\
        & \texttt{Input: Consider the music of <Music> and its paired image of <Image> that visually represents a key moment of the music.} \\
        & \texttt{Instruction: Provide a description of how the music's rhythm, tempo, and tonal qualities are visually represented in the image, combining insights from both the music and the image captions.} \\
        & \texttt{Output: The music is characterized by a slow tempo with a calm, serene melody, primarily featuring piano and soft strings. The image complements this mood, showing a peaceful sunset over a calm ocean. The soft, gentle waves in the image mirror the steady downbeats of the music, while the warm color palette in the image reflects the key of the music, which is in A major, creating a sense of tranquility. Together, they evoke a feeling of peace and reflection.} \\
        \hline
        \multirow{14}{*}{Output} & \texttt{Input: Consider the music of <Music>; and its paired image of <Image>; or video of <Video>; that visually represents a dark and intense atmosphere.} \\
        & \texttt{Instruction: Describe how the music's slow, heavy metal style and the visuals of a man playing a guitar are reflected in the image or video.} \\
        & \texttt{Output: The music is a slow, heavy metal song with a dark and intense mood. It features electric guitar, bass guitar, drums, and powerful vocals. The tempo is slow and the rhythm is steady, with a heavy, distorted guitar sound and a solid foundation provided by the bass guitar and drums. The vocals are deep and powerful, emphasizing the lower register. This creates a heavy and intense sound. The image or video shows a man playing a guitar, holding it, and sitting on a table. The visuals are simple but effective, focusing on the man and his instrument. The dark and intense atmosphere of the music is reflected in the visuals, capturing the mood and energy of the song. Together, the music and visuals create a cohesive experience that immerses the audience in a dark and intense musical journey.} \\
    \bottomrule
    \end{tabular}
    }
    \caption{Prompting GPT-4o mini to generate triplets of input, instruction, and output for Music4way-Any2T. The instruction represents the system description used when prompting GPT-4o mini.}
    \label{tab:any2t}
\end{table*}

\begin{table*}[t]
    \centering
    \resizebox{\linewidth}{!}{
    \begin{tabular}{l|p{\linewidth}}
    \toprule
        \multicolumn{2}{c}{Example 1} \\
        \hline
        \multirow{3}{*}{Input} & \texttt{The music from <Music> plays alongside a captivating scene of <Video> where a person is sitting and showcasing their musical talent with a string instrument, which is also reflected in <Image>.} \\
        \hline
        \multirow{3}{*}{Instruction} & \texttt{Analyze how the mood and emotion conveyed through the music interact with the visual elements presented in the video and image, focusing on the cohesive atmosphere they create together.} \\
        \hline
        \multirow{10}{*}{Output} & \texttt{The music is characterized by a slow tempo and intense instrumental arrangement, prominently featuring the guitar which serves as the main tune. The video depicts a person deeply focused while playing a string instrument, highlighting their connection with the music. This is mirrored in the image of the man playing the guitar, showcasing the intimacy of this musical expression. The harmonious blend of the electric guitar enhances the emotional depth, adding layers to the music that evoke feelings of intensity. The consistent downbeats and the key of D minor resonate with the serious tone of the visuals. Together, the visual components and the auditory experience create a mesmerising atmosphere, pulling the viewer into a world where music and emotion intertwine seamlessly.} \\
    \toprule
        \multicolumn{2}{c}{Example 2} \\
        \hline
        \multirow{3}{*}{Input} & \texttt{Consider the music of <Music>, which is a slow and mellow instrumental with a melancholic mood, alongside the paired video of <Video> where a man is playing a song on a guitar.} \\
        \hline
        \multirow{3}{*}{Instruction} & \texttt{Explain how the guitar performance in the video enhances the melancholic atmosphere of the music, and describe how both elements work together to create an emotional experience for the viewer.} \\
        \hline
        \multirow{10}{*}{Output} & \texttt{The music, characterized by its slow tempo and melancholic mood, beautifully complements the visuals of the man playing the guitar in the video. As he strums the instrument on the table, the gentle, steady rhythm of the music aligns with his movements, emphasizing the emotional weight of each chord he plays. The use of major and minor chords, particularly in the F\# major scale, enhances the bittersweet feelings conveyed by the music. The visuals of the man fully engrossed in his performance highlight the introspective nature of the piece, creating a profound connection between the audio and the visual. Together, they evoke a powerful sense of sadness and reflection, inviting the viewer to immerse themselves in this poignant moment.} \\
    \toprule
        \multicolumn{2}{c}{Example 3} \\
        \hline
        \multirow{3}{*}{Input} & \texttt{Consider the music of <Music> that is fast-paced and energetic, while viewing the video of <Video> showing a young man playing the guitar, alongside the image of <Image> depicting the same scene.} \\
        \hline
        \multirow{3}{*}{Instruction} & \texttt{Describe how the fast-paced rhythm and energetic tone of the music interact with the visuals of the video and image, focusing on the emotions and energy conveyed by both the audio and visual components.} \\
        \hline
        \multirow{11}{*}{Output} & \texttt{The fast-paced music, characterized by a strong beat and an upbeat tempo, perfectly complements the visuals of the young man playing guitar in both the video and the image. The upbeat rhythm creates a vibrant atmosphere, as the energetic strumming of the guitar aligns with the driving drums and pulsating bass. The 4/4 time signature of the music further emphasizes the lively performance, with each downbeat coinciding with the man's movements. The use of E major and A major chords adds a bright and happy tonal quality that is visually represented in both the video and image through the man's engaging expression and dynamic posture, conveying a sense of excitement and intensity that radiates through the screen. Together, the music and visuals fuse seamlessly, creating an exhilarating experience that captures the essence of a lively performance.} \\
    \bottomrule
    \end{tabular}
    }
    \caption{More examples of Music4way-Any2T for evaluation. ``<Music>'', ``<Image>'', and ``<Video>'' represent the paths to the respective files.}
    \label{tab:any2tdata}
\end{table*}

\begin{table*}[t]
    \centering
    \resizebox{\linewidth}{!}{
    \begin{tabular}{lrrrrrrrr}
    \toprule
    \textbf{Model} & \textbf{BLEU-1} & \textbf{BLEU} & \textbf{ROUGE-P} & \textbf{ROUGE-R} & \textbf{ROUGE-F1} & \textbf{BERT-P} & \textbf{BERT-R} & \textbf{BERT-F1} \\
    \toprule
    \textbf{SALMONN}~\cite{DBLP:conf/iclr/TangYSC000M024}$^\dagger$ & 28.7 & 13.1 & 45.7 & 37.2 & 35.4 & 90.8 & 89.9 & 90.3 \\
    \textbf{MU-LLaMA}~\cite{DBLP:conf/icassp/LiuHSS24}$^\dagger$ & 29.7 & 10.9 & 30.8 & \textbf{46.4} & 33.1 & 89.8 & 90.0 & 89.9 \\
    \textbf{OpenMU}~\cite{DBLP:journals/corr/abs-2410-15573}$^\dagger$ & 24.5 & 6.1 & 20.8 & \textbf{44.8} & 25.5 & 86.9 & 90.5 & 88.6 \\
    \textbf{NExT-GPT}~\cite{DBLP:conf/icml/Wu0Q0C24} & 23.3 & 7.6 & 26.2 & 38.4 & 26.0 & 86.7 & 88.5 & 87.6 \\
    \textbf{M$^2$UGen}~\cite{DBLP:journals/corr/abs-2311-11255}$^\dagger$ & 29.1 & 13.1 & \textbf{52.7} & 35.0 & 37.9 & 91.7 & 89.3 & 90.5 \\
    \textbf{NExT-GPT w/ M$^2$UGen}$^\dagger$ & 34.0 & 14.3 & 49.1 & 38.3 & 39.6 & 92.1 & 90.5 & 91.2 \\
    \textbf{NExT-GPT w/ Music4way}$^\dagger$ & 34.1 & 13.8 & 48.2 & 37.9 & 39.2 & 92.1 & 90.5 & 91.3 \\
    \rowcolor{gray!20}
    \textbf{DeepResonance-$\alpha$ (ours)}$^\dagger$ & \textbf{35.1} & \textbf{15.1} & 51.0 & 38.6 & \textbf{40.8} & \textbf{92.5} & \textbf{90.7} & \textbf{91.6} \\
    \rowcolor{gray!20}
    \textbf{DeepResonance-$\beta$ (ours)}$^\dagger$ & \textbf{35.6} & \textbf{15.3} & \textbf{51.3} & 39.0 & \textbf{41.1} & \textbf{92.5} & \textbf{90.8} & \textbf{91.6} \\
    \hline
    \textbf{NExT-GPT-Mistral w/ Music4way}$^\dagger$ & 33.9 & 14.1 & 50.3 & 37.2 & 39.7 & 92.2 & 90.3 & 91.2 \\
    \rowcolor{gray!20}
    \textbf{DeepResonance-$\alpha$-Mistral (ours)}$^\dagger$ & \textbf{35.0} & \textbf{14.7} & 50.0 & \textbf{38.5} & \textbf{40.4} & \textbf{92.3} & \textbf{90.6} & \textbf{91.4} \\
    \rowcolor{gray!20}
    \textbf{DeepResonance-$\beta$-Mistral (ours)}$^\dagger$ & 
    \textbf{34.6} & \textbf{14.6} & \textbf{50.9} & \textbf{38.0} & \textbf{40.5} & \textbf{92.4} & \textbf{90.5} & \textbf{91.4} \\
    \hline
    \textbf{NExT-GPT-LLaMA-3 w/ Music4way}$^\dagger$ & 32.3 & 13.1 & 50.3 & 36.0 & 38.8 & 92.3 & 90.2 & 91.2 \\
    \rowcolor{gray!20}
    \textbf{DeepResonance-$\alpha$-LLaMA-3 (ours)}$^\dagger$ & \textbf{33.9} & \textbf{13.9} & 49.5 & \textbf{37.6} & \textbf{39.5} & 92.1 & \textbf{90.4} & 91.2 \\
    \rowcolor{gray!20}
    \textbf{DeepResonance-$\beta$-LLaMA-3 (ours)}$^\dagger$ & \textbf{33.3} & \textbf{13.8} & \textbf{50.4} & \textbf{36.8} & \textbf{39.4} & 92.2 & 90.2 & 91.2 \\
    \bottomrule
    \end{tabular}
    }
    \caption{\textbf{Results on MusicQA.} The top two performances in the first block are highlighted in \textbf{bold}. Scores where Mistral or LLaMA-3 based models surpass their respective baselines are shown in \textbf{bold}. ``$^\dagger$'' indicates supervised evaluation settings.}
    \label{tab:musicqa}
\end{table*}

\begin{table*}[t]
    \centering
    \resizebox{\linewidth}{!}{
    \begin{tabular}{lrrrrrrrr}
    \toprule
    \textbf{Model} & \textbf{BLEU-1} & \textbf{BLEU} & \textbf{ROUGE-P} & \textbf{ROUGE-R} & \textbf{ROUGE-F1} & \textbf{BERT-P} & \textbf{BERT-R} & \textbf{BERT-F1} \\
    \toprule
    \textbf{SALMONN}~\cite{DBLP:conf/iclr/TangYSC000M024}$^\dagger$ & 19.7 & 0.9 & 23.6 & 21.5 & 19.1 & \textbf{87.6} & 86.3 & 86.9 \\
    \textbf{MU-LLaMA}~\cite{DBLP:conf/icassp/LiuHSS24}$^\dagger$ & $^*$9.6 & $^*$0.2 & $^*$32.0 & $^*$12.0 & $^*$16.2 & $^*$88.7 & $^*$85.1 & $^*$86.8 \\
    \textbf{OpenMU}~\cite{DBLP:journals/corr/abs-2410-15573}$^\dagger$ & 23.9 & 1.2 & 18.8 & \textbf{22.8} & 19.4 & 86.1 & \textbf{87.2} & 86.6 \\
    \textbf{MusiLingo sft. w/ MusicCaps}~\cite{deng-etal-2024-musilingo}$^\dagger$ & -- & -- & -- & -- & \textbf{21.7} & -- & -- & 86.8 \\
    \textbf{NExT-GPT}~\cite{DBLP:conf/icml/Wu0Q0C24} & 16.5 & 0.1 & 15.3 & 16.4 & 14.0 & 84.3 & 83.8 & 84.0 \\
    \textbf{M$^2$UGen}~\cite{DBLP:journals/corr/abs-2311-11255}$^\dagger$ & $^*$14.4 & $^*$0.4 & $^*$26.1 & $^*$14.5 & $^*$16.4 & $^*$87.8 & $^*$85.4 & $^*$86.5 \\
    \textbf{NExT-GPT w/ M$^2$UGen} & 12.4 & 0.1 & \textbf{28.2} & 13.2 & 16.1 & \textbf{88.6} & 85.3 & 86.9 \\
    \textbf{NExT-GPT w/ Music4way}$^\dagger$ & 25.0 & 2.7 & 23.0 & 20.6 & 21.0 & \textbf{87.6} & 87.0 & 87.2 \\
    \rowcolor{gray!20}
    \textbf{DeepResonance-$\alpha$ (ours)}$^\dagger$ & \textbf{26.0} & \textbf{3.0} & 23.4 & \textbf{21.8} & \textbf{21.6} & \textbf{87.6} & \textbf{87.1} & \textbf{87.3} \\
    \rowcolor{gray!20}
    \textbf{DeepResonance-$\beta$ (ours)}$^\dagger$ & \textbf{25.8} & \textbf{2.8} & \textbf{23.6} & 21.4 & \textbf{21.6} & 87.5 & \textbf{87.1} & \textbf{87.3} \\
    \hline
    \textbf{NExT-GPT-Mistral w/ Music4way}$^\dagger$ & 25.3 & 2.7 &  22.6 & 21.1 & 20.8 & 87.4 & 86.9 & 87.1 \\
    \rowcolor{gray!20}
    \textbf{DeepResonance-$\alpha$-Mistral (ours)}$^\dagger$ & \textbf{25.8} & \textbf{2.8} &  \textbf{23.1} & \textbf{21.6} & \textbf{21.3} & \textbf{87.5} & \textbf{87.0} & \textbf{87.2} \\
    \rowcolor{gray!20}
    \textbf{DeepResonance-$\beta$-Mistral (ours)}$^\dagger$ & \textbf{25.6} & 2.6 &  22.5 & \textbf{21.4} & \textbf{21.0} & 87.4 & \textbf{87.0} & \textbf{87.2} \\
    \hline
    \textbf{NExT-GPT-LLaMA-3 w/ Music4way}$^\dagger$ & 25.4 & 2.7 &  22.8 & 21.2 & 21.0 & 87.4 & 86.8 & 87.1 \\
    \rowcolor{gray!20}
    \textbf{DeepResonance-$\alpha$-LLaMA-3 (ours)}$^\dagger$ & \textbf{26.0} & 2.7 &  \textbf{23.0} & \textbf{21.6} & \textbf{21.3} & \textbf{87.5} & \textbf{87.0} & \textbf{87.2} \\
    \rowcolor{gray!20}
    \textbf{DeepResonance-$\beta$-LLaMA-3 (ours)}$^\dagger$ & \textbf{25.8} & \textbf{2.8} &  \textbf{23.1} & \textbf{21.6} & \textbf{21.3} & \textbf{87.5} & \textbf{87.0} & \textbf{87.2} \\
    \bottomrule
    \end{tabular}
    }
    \caption{\textbf{Results on MusicCaps.} The top two performances in the first block are highlighted in \textbf{bold}. Scores where Mistral or LLaMA-3 based models surpass their respective baselines are shown in \textbf{bold}. ``*'' denotes results that should be interpreted with caution, as the test data was included in the corresponding model's training set. ``$^\dagger$'' indicates supervised evaluation settings.}
    \label{tab:musiccaps}
\end{table*}

\begin{table*}[t!]
    \centering
    \resizebox{\linewidth}{!}{
    \begin{tabular}{lrrrrrrrr}
    \toprule
    \textbf{Model} & \textbf{BLEU-1} & \textbf{BLEU} & \textbf{ROUGE-P} & \textbf{ROUGE-R} & \textbf{ROUGE-F1} & \textbf{BERT-P} & \textbf{BERT-R} & \textbf{BERT-F1} \\
    \toprule
    \textbf{SALMONN}~\cite{DBLP:conf/iclr/TangYSC000M024} & 19.1 & 0.8 & 15.7 & 39.8 & 20.0 & 85.5 & 88.6 & 87.0 \\
    \textbf{MU-LLaMA}~\cite{DBLP:conf/icassp/LiuHSS24} & 15.1 & 0.4 & 52.1 & 20.9 & 27.6 & 90.1 & 86.6 & 88.3 \\
    \textbf{NExT-GPT}~\cite{DBLP:conf/icml/Wu0Q0C24} & 16.6 & 0.2 & 12.4 & 37.0 & 17.2 & 85.3 & 88.2 & 86.7 \\
    \textbf{M$^2$UGen}~\cite{DBLP:journals/corr/abs-2311-11255}$^\dagger$ & $^*$13.1 & $^*$0.0 & $^*$52.8 & $^*$19.5 & $^*$26.0 & $^*$89.6 & $^*$85.8 & $^*$87.6 \\
    \textbf{NExT-GPT w/ M$^2$UGen}$^\dagger$ & $^*$23.2 & $^*$6.7 & $^*$60.8 & $^*$29.7 & $^*$36.7 & $^*$93.1 & $^*$89.6 & $^*$91.3 \\
    \textbf{NExT-GPT w/ Music4way}$^\dagger$ & 39.1 & 18.4 & 55.5 & 46.9 & 46.8 & 91.7 & \textbf{92.5} & 93.0 \\
    \rowcolor{gray!20}
    \textbf{DeepResonance-$\alpha$ (ours)}$^\dagger$ & \textbf{40.9} & \textbf{19.9} & \textbf{57.8} & \textbf{47.5} & \textbf{48.4} & \textbf{94.0} & \textbf{92.6} & \textbf{93.3} \\
    \rowcolor{gray!20}
    \textbf{DeepResonance-$\beta$ (ours)}$^\dagger$ & \textbf{39.9} & \textbf{19.3} & \textbf{57.3} & \textbf{47.3} & \textbf{47.8} & \textbf{93.9} & \textbf{92.5} & \textbf{93.2} \\
    \hline
    \textbf{NExT-GPT-Mistral w/ Music4way}$^\dagger$ & 37.9 & 16.9 & 56.9 & 44.5 & 46.2 & 93.8 & 92.2 & 93.0 \\
    \rowcolor{gray!20}
    \textbf{DeepResonance-$\alpha$-Mistral (ours)}$^\dagger$ & \textbf{39.1} & \textbf{18.6} & \textbf{59.4} & \textbf{45.1} & \textbf{47.9} & \textbf{94.2} & 92.2 & \textbf{93.2} \\
    \rowcolor{gray!20}
    \textbf{DeepResonance-$\beta$-Mistral (ours)}$^\dagger$ & \textbf{39.7} & \textbf{18.9} & \textbf{58.4} & \textbf{45.7} & \textbf{47.8} & \textbf{94.0} & \textbf{92.4} & \textbf{93.2} \\
    \hline
    \textbf{NExT-GPT-LLaMA-3 w/ Music4way}$^\dagger$ & 37.9 & 16.9 & 56.9 & 44.5 & 46.2 & 93.8 & 92.2 & 93.0 \\
    \rowcolor{gray!20}
    \textbf{DeepResonance-$\alpha$-LLaMA-3 (ours)}$^\dagger$ & \textbf{39.5} & \textbf{19.2} & \textbf{60.3} & \textbf{45.4} & \textbf{48.4} & \textbf{94.4} & \textbf{92.3} & \textbf{93.3} \\
    \rowcolor{gray!20}
    \textbf{DeepResonance-$\beta$-LLaMA-3 (ours)}$^\dagger$ & \textbf{39.7} & \textbf{19.2} & \textbf{60.3} & \textbf{45.2} & \textbf{48.4} & \textbf{94.4} & \textbf{92.3} & \textbf{93.3} \\
    \bottomrule
    \end{tabular}
    }
    \caption{\textbf{Results on Music4way-MusicCaps.} The top two performances in the first block are highlighted in \textbf{bold}. Scores where Mistral or LLaMA-3 based models surpass their respective baselines are shown in \textbf{bold}. ``*'' denotes results that should be interpreted with caution, as the test data was included in the corresponding model's training set. ``$^\dagger$'' indicates supervised evaluation settings.}
    \label{tab:music4way-musiccaps}
\end{table*}

\begin{table*}[t]
    \centering
    \resizebox{\linewidth}{!}{
    \begin{tabular}{lrrrrrrrr}
    \toprule
    \textbf{Model} & \textbf{BLEU-1} & \textbf{BLEU} & \textbf{ROUGE-P} & \textbf{ROUGE-R} & \textbf{ROUGE-F1} & \textbf{BERT-P} & \textbf{BERT-R} & \textbf{BERT-F1} \\
    \toprule
    \textbf{NExT-GPT}~\cite{DBLP:conf/icml/Wu0Q0C24} & 26.7 & 1.2 & 26.2 & 19.7 & 21.3 & 85.5 & 85.0 & 85.2 \\
    \textbf{M$^2$UGen}~\cite{DBLP:journals/corr/abs-2311-11255} & $^*$31.7 & $^*$4.1 & $^*$29.5 & $^*$26.5 & $^*$26.4 & $^*$87.7 & $^*$86.6 & $^*$87.1 \\
    \textbf{NExT-GPT w/ M$^2$UGen} & $^*$33.8 & $^*$4.1 & $^*$35.0 & $^*$23.9 & $^*$27.3 & $^*$89.0 & $^*$87.3 & $^*$88.1 \\
    \textbf{NExT-GPT w/ Music4way} & 24.2 & 1.7 & 25.1 & 23.6 & 22.0 & 85.7 & 85.3 & 85.4 \\
    \rowcolor{gray!20}
    \textbf{DeepResonance-$\alpha$ (ours)} & \textbf{48.7} & \textbf{16.4} & \textbf{37.1} & \textbf{36.8} & \textbf{36.2} & \textbf{90.3} & \textbf{90.2} & \textbf{90.2} \\
    \rowcolor{gray!20}
    \textbf{DeepResonance-$\beta$ (ours)} & \textbf{49.2} & \textbf{17.2} & \textbf{36.8} & \textbf{38.3} & \textbf{36.8} & \textbf{90.1} & \textbf{90.3} & \textbf{90.2} \\
    \hline
    \textbf{NExT-GPT-Mistral w/ Music4way} & 8.6 & 0.8 & 45.6 & 12.6 & 18.7 & 89.6 & 85.2 & 87.3\\
    \rowcolor{gray!20}
    \textbf{DeepResonance-$\alpha$-Mistral (ours)} & \textbf{48.9} & \textbf{16.6} & 37.8 & \textbf{36.9} & \textbf{36.7} & \textbf{90.4} & \textbf{90.2} & \textbf{90.3} \\
    \rowcolor{gray!20}
    \textbf{DeepResonance-$\beta$-Mistral (ours)} & \textbf{49.8} & \textbf{17.5} & 37.5 & \textbf{38.3} & \textbf{37.2} & \textbf{90.3} & \textbf{90.4} & \textbf{90.4} \\
    \hline
    \textbf{NExT-GPT-LLaMA-3 w/ Music4way} & 5.7 & 0.2 & 37.7 & 10.4 & 15.0 & 86.9 & 83.1 & 84.9 \\
    \rowcolor{gray!20}
    \textbf{DeepResonance-$\alpha$-LLaMA-3 (ours)} & \textbf{48.9} & \textbf{16.5} & \textbf{37.5} & \textbf{36.8} & \textbf{36.4} & \textbf{90.3} & \textbf{90.2} & \textbf{90.3} \\
    \rowcolor{gray!20}
    \textbf{DeepResonance-$\beta$-LLaMA-3 (ours)} & \textbf{49.4} & \textbf{17.3} & \textbf{37.3} & \textbf{38.2} & \textbf{37.0} & \textbf{90.3} & \textbf{90.4} & \textbf{90.3} \\
    \bottomrule
    \end{tabular}
    }
    \caption{\textbf{Results on Music4way-MI2T.} The top two performances in the first block are highlighted in \textbf{bold}. Scores where Mistral or LLaMA-3 based models surpass their respective baselines are shown in \textbf{bold}. ``*'' denotes results that should be interpreted with caution, as the test data was included in the corresponding model's training set.}
    \label{tab:music4way-mi2t}
\end{table*}

\begin{table*}[t]
    \centering
    \resizebox{\linewidth}{!}{
    \begin{tabular}{lrrrrrrrr}
    \toprule
    \textbf{Model} & \textbf{BLEU-1} & \textbf{BLEU} & \textbf{ROUGE-P} & \textbf{ROUGE-R} & \textbf{ROUGE-F1} & \textbf{BERT-P} & \textbf{BERT-R} & \textbf{BERT-F1} \\
    \toprule
    \textbf{NExT-GPT}~\cite{DBLP:conf/icml/Wu0Q0C24} & 26.5 & 1.1 & 25.6 & 19.6 & 21.0 & 85.1 & 84.6 & 84.8 \\
    \textbf{M$^2$UGen}~\cite{DBLP:journals/corr/abs-2311-11255} & $^*$31.7 & $^*$4.3 & $^*$27.9 & $^*$26.5 & $^*$25.9 & $^*$87.2 & $^*$86.4 & $^*$86.8 \\
    \textbf{NExT-GPT w/ M$^2$UGen} & $^*$34.6 & $^*$4.0 & $^*$34.8 & $^*$24.2 & $^*$27.3 & $^*$88.9 & $^*$87.4 & $^*$88.1 \\
    \textbf{NExT-GPT w/ Music4way} & 25.0 & 1.8 & 25.4 & 24.2 & 22.5 & 85.7 & 85.4 & 85.5 \\
    \rowcolor{gray!20}
    \textbf{DeepResonance-$\alpha$ (ours)} & \textbf{48.9} & \textbf{16.6} & \textbf{37.4} & \textbf{37.0} & \textbf{36.5} & \textbf{90.3} & \textbf{90.2} & \textbf{90.2} \\
    \rowcolor{gray!20}
    \textbf{DeepResonance-$\beta$ (ours)} & \textbf{49.0} & \textbf{17.2} & \textbf{36.7} & \textbf{38.4} & \textbf{36.8} & \textbf{90.3} & \textbf{90.3} & \textbf{90.3} \\
    \hline
    \textbf{NExT-GPT-Mistral w/ Music4way} & 9.0 & 0.9 & 46.1 & 12.8 & 19.0 & 89.7 & 85.3 & 87.4 \\
    \rowcolor{gray!20}
    \textbf{DeepResonance-$\alpha$-Mistral (ours)} & \textbf{48.7} & \textbf{16.4} & 37.5 & \textbf{36.7} & \textbf{36.4} & \textbf{90.4} & \textbf{90.2} & \textbf{90.3} \\
    \rowcolor{gray!20}
    \textbf{DeepResonance-$\beta$-Mistral (ours)} & \textbf{49.7} & \textbf{17.5} & 37.5 & \textbf{38.3} & \textbf{37.2} & \textbf{90.3} & \textbf{90.4} & \textbf{90.4} \\
    \hline
    \textbf{NExT-GPT-LLaMA-3 w/ Music4way} & 4.9 & 0.2 & 38.5 & 10.5 & 14.9 & 86.8 & 83.1 & 84.9 \\
    \rowcolor{gray!20}
    \textbf{DeepResonance-$\alpha$-LLaMA-3 (ours)} & \textbf{49.0} & \textbf{16.7} & \textbf{37.6} & \textbf{36.8} & \textbf{36.5} & \textbf{90.4} & \textbf{90.2} & \textbf{90.3} \\
    \rowcolor{gray!20}
    \textbf{DeepResonance-$\beta$-LLaMA-3 (ours)} & \textbf{49.5} & \textbf{17.5} & \textbf{37.2} & \textbf{38.2} & \textbf{36.9} & \textbf{90.2} & \textbf{90.4} & \textbf{90.3} \\
    \bottomrule
    \end{tabular}
    }
    \caption{\textbf{Results on Music4way-MV2T.} The top two performances in the first block are highlighted in \textbf{bold}. Scores where Mistral or LLaMA-3 based models surpass their respective baselines are shown in \textbf{bold}. ``*'' denotes results that should be interpreted with caution, as the test data was included in the corresponding model's training set.}
    \label{tab:music4way-mv2t}
\end{table*}

\begin{table*}[t!]
    \centering
    \resizebox{\linewidth}{!}{
    \begin{tabular}{lrrrrrrrr}
    \toprule
    \textbf{Model} & \textbf{BLEU-1} & \textbf{BLEU} & \textbf{ROUGE-P} & \textbf{ROUGE-R} & \textbf{ROUGE-F1} & \textbf{BERT-P} & \textbf{BERT-R} & \textbf{BERT-F1} \\
    \toprule
    \textbf{NExT-GPT}~\cite{DBLP:conf/icml/Wu0Q0C24} & 25.4 & 3.4 & 33.5 & 19.8 & 23.4 & 87.4 & 85.9 & 86.6 \\
    \textbf{M$^2$UGen}~\cite{DBLP:journals/corr/abs-2311-11255} & $^*$20.8 & $^*$2.5 & $^*$34.4 & $^*$17.1 & $^*$21.5 & $^*$88.9 & $^*$85.8 & $^*$87.3 \\
    \textbf{NExT-GPT w/ M$^2$UGen} & $^*$26.3 & $^*$4.8 & $^*$41.7 & $^*$22.2 & $^*$28.5 & $^*$91.1 & $^*$87.7 & $^*$89.4 \\
    \textbf{NExT-GPT w/ Music4way} & 29.4 & 2.8 & \textbf{35.9} & 22.7 & 27.0 & 89.7 & 87.1 & 88.4 \\
    \rowcolor{gray!20}
    \textbf{DeepResonance-$\alpha$ (ours)} & \textbf{37.2} & \textbf{6.0} & \textbf{36.0} & \textbf{25.7} & \textbf{29.6} & \textbf{90.8} & \textbf{88.3} & \textbf{89.5} \\
    \rowcolor{gray!20}
    \textbf{DeepResonance-$\beta$ (ours)} & \textbf{33.5} & \textbf{3.9} & 34.0 & \textbf{23.9} & \textbf{27.4} & \textbf{90.0} & \textbf{87.5} & \textbf{88.7} \\
    \hline
    \textbf{NExT-GPT-Mistral w/ Music4way} & 17.4 & 1.8 &  41.8 & 17.8 & 24.0 & 90.0 & 86.2 & 88.0 \\
    \rowcolor{gray!20}
    \textbf{DeepResonance-$\alpha$-Mistral (ours)} & \textbf{37.2} & \textbf{6.1} & 35.4 & \textbf{26.2} & \textbf{29.4} & \textbf{90.6} & \textbf{88.2} & \textbf{89.4} \\
    \rowcolor{gray!20}
    \textbf{DeepResonance-$\beta$-Mistral (ours)} & \textbf{34.8} & \textbf{3.9} & 32.3 & \textbf{24.2} & \textbf{26.8} & \textbf{89.6} & \textbf{87.5} & \textbf{88.6} \\
    \hline
    \textbf{NExT-GPT-LLaMA-3 w/ Music4way} & 25.7 & 2.9 &  35.9 & 21.1 & 25.4 & 89.1 & 86.4 & 87.7 \\
    \rowcolor{gray!20}
    \textbf{DeepResonance-$\alpha$-LLaMA-3 (ours)} & \textbf{34.8} & \textbf{5.7} & \textbf{37.2} & \textbf{25.0} & \textbf{29.4} & \textbf{90.9} & \textbf{88.2} & \textbf{89.5} \\
    \rowcolor{gray!20}
    \textbf{DeepResonance-$\beta$-LLaMA-3 (ours)} & \textbf{35.4} & \textbf{4.5} & 33.9 & \textbf{24.8} & \textbf{28.0} & \textbf{90.1} & \textbf{87.8} & \textbf{88.9} \\
    \bottomrule
    \end{tabular}
    }
    \caption{\textbf{Results on Music4way-Any2T.} The top two performances in the first block are highlighted in \textbf{bold}. Scores where Mistral or LLaMA-3 based models surpass their respective baselines are shown in \textbf{bold}. ``*'' denotes results that should be interpreted with caution, as the test data was included in the corresponding model's training set.}
    \label{tab:music4way-any2t}
\end{table*}

\begin{table*}[t]
    \centering
    \resizebox{\linewidth}{!}{
    \begin{tabular}{lrrrrrrrr}
    \toprule
    \textbf{Model} & \textbf{BLEU-1} & \textbf{BLEU} & \textbf{ROUGE-P} & \textbf{ROUGE-R} & \textbf{ROUGE-F1} & \textbf{BERT-P} & \textbf{BERT-R} & \textbf{BERT-F1} \\
    \toprule
    \textbf{NExT-GPT}~\cite{DBLP:conf/icml/Wu0Q0C24} & \textbf{19.3} & 1.0 & 22.2 & 16.7 & 16.9 & 85.0 & 84.2 & 84.6 \\
    \textbf{NExT-GPT w/ M$^2$UGen} & 15.4 & 2.9 & \textbf{38.6} & 16.6 & 21.7 & 88.9 & 86.3 & 87.5 \\
    \textbf{NExT-GPT w/ Music4way} & 16.3 & 2.6 & 37.2 & 17.2 & 22.5 & \textbf{89.6} & 86.9 & \textbf{88.2} \\
    \rowcolor{gray!20}
    \textbf{DeepResonance-$\alpha$ (ours)} & 19.2 & \textbf{3.1} & 35.8 & \textbf{18.7} & \textbf{23.0} & 89.4 & \textbf{87.0} & \textbf{88.2} \\
    \rowcolor{gray!20}
    \textbf{DeepResonance-$\beta$ (ours)} & \textbf{19.5} & \textbf{4.5} & \textbf{39.2} & \textbf{20.1} & \textbf{25.2} & \textbf{90.1} & \textbf{87.5} & \textbf{88.8} \\
    \hline
    \textit{OpenMU (supervised performance upper bound)} & \textit{47.2} & \textit{18.5} & \textit{35.7} & \textit{41.2} & \textit{37.5} & \textit{91.1} & \textit{91.9} & \textit{91.5} \\
    \bottomrule
    \end{tabular}
    }
    \caption{\textbf{Results on GTZAN.} The top two performances are highlighted in \textbf{bold}.}
    \label{tab:gtzan}
\end{table*}

\begin{table*}[t]
    \centering
    \resizebox{\linewidth}{!}{
    \begin{tabular}{lrrrrrrrr}
    \toprule
    \textbf{Model} & \textbf{BLEU-1} & \textbf{BLEU} & \textbf{ROUGE-P} & \textbf{ROUGE-R} & \textbf{ROUGE-F1} & \textbf{BERT-P} & \textbf{BERT-R} & \textbf{BERT-F1} \\
    \toprule
    \textbf{NExT-GPT}~\cite{DBLP:conf/icml/Wu0Q0C24} & \textbf{19.5} & 0.0 & 23.6 & 10.1 & 13.5 & 83.7 & 83.2 & 83.4 \\
    \textbf{NExT-GPT w/ M$^2$UGen} & 4.3 & \textbf{0.2} & \textbf{42.0} & 9.4 & 14.1 & \textbf{88.0} & 84.8 & 85.8 \\
    \textbf{NExT-GPT w/ Music4way} & 13.7 & \textbf{0.1} & 24.8 & 12.2 & 15.8 & 86.9 & \textbf{85.2} & 86.0 \\
    \rowcolor{gray!20}
    \textbf{DeepResonance-$\alpha$ (ours)} & 14.6 & \textbf{0.1} & 24.4 & \textbf{12.5} & \textbf{15.9} & \textbf{87.0} & \textbf{85.2} & \textbf{86.1} \\
    \rowcolor{gray!20}
    \textbf{DeepResonance-$\beta$ (ours)} & \textbf{14.9} & \textbf{0.1} & \textbf{25.1} & \textbf{12.7} & \textbf{16.4} & \textbf{87.0} & \textbf{85.6} & \textbf{86.2} \\
    \hline
    \textit{OpenMU (supervised performance upper bound)} & \textit{52.4} & \textit{22.0} & \textit{35.8} & \textit{44.4} & \textit{39.3} & \textit{91.5} & \textit{92.4} & \textit{92.0} \\
    \bottomrule
    \end{tabular}
    }
    \caption{\textbf{Results on MusicNet.} The top two performances are highlighted in \textbf{bold}.}
    \label{tab:musicnet}
\end{table*}

\begin{table*}[t!]
    \centering
    \resizebox{\linewidth}{!}{
    \begin{tabular}{lrrrrrrrr}
    \toprule
    \textbf{Model} & \textbf{BLEU-1} & \textbf{BLEU} & \textbf{ROUGE-P} & \textbf{ROUGE-R} & \textbf{ROUGE-F1} & \textbf{BERT-P} & \textbf{BERT-R} & \textbf{BERT-F1} \\
    \toprule
    \textbf{NExT-GPT}~\cite{DBLP:conf/icml/Wu0Q0C24} & \textbf{13.9} & 0.1 & 23.7 & \textbf{12.2} & 15.4 & 85.7 & \textbf{84.2} & 84.9 \\
    \textbf{NExT-GPT w/ M$^2$UGen} & 7.1 & 0.1 & 33.0 & 10.2 & 15.3 & 87.6 & 83.6 & 85.6 \\
    \textbf{NExT-GPT w/ Music4way} & 7.3 & \textbf{0.2} & 32.7 & 10.4 & 15.5 & 87.8 & 83.8 & 85.8 \\
    \rowcolor{gray!20}
    \textbf{DeepResonance-$\alpha$ (ours)} & 8.6 & \textbf{0.2} & \textbf{35.9} & 10.5 & \textbf{15.8} & \textbf{88.4} & 83.8 & \textbf{86.0} \\
    \rowcolor{gray!20}
    \textbf{DeepResonance-$\beta$ (ours)} & \textbf{9.4} & \textbf{0.2} & \textbf{33.6} & \textbf{11.4} & \textbf{16.5} & \textbf{88.4} & \textbf{84.0} & \textbf{86.1} \\
    \hline
    \textit{MusiLingo (supervised performance upper bound)} & \textit{45.0} & -- & -- & -- & \textit{22.9} & -- & -- & \textit{86.1} \\
    \bottomrule
    \end{tabular}
    }
    \caption{\textbf{Results on MusicInstruct-long.} The top two performances are highlighted in \textbf{bold}.}
    \label{tab:musicinstructlong}
\end{table*}

\begin{table*}[t!]
    \centering
    \resizebox{\linewidth}{!}{
    \begin{tabular}{lrrrrrrrr}
    \toprule
    \textbf{Model} & \textbf{BLEU-1} & \textbf{BLEU} & \textbf{ROUGE-P} & \textbf{ROUGE-R} & \textbf{ROUGE-F1} & \textbf{BERT-P} & \textbf{BERT-R} & \textbf{BERT-F1} \\
    \toprule
    \textbf{NExT-GPT}~\cite{DBLP:conf/icml/Wu0Q0C24} & 28.9 & 10.3 & 28.6 & \textbf{50.1} & 32.1 & 88.5 & 90.9 & 89.6 \\
    \textbf{NExT-GPT w/ M$^2$UGen} & 41.5 & 15.5 & 47.3 & 46.8 & 44.6 & 92.5 & 91.8 & 92.1 \\
    \textbf{NExT-GPT w/ Music4way} & 42.9 & 16.4 & 47.8 & 46.5 & 44.9 & 92.7 & 91.9 & 92.3 \\
    \rowcolor{gray!20}
    \textbf{DeepResonance-$\alpha$ (ours)} & \textbf{43.0} & \textbf{16.8} & \textbf{48.0} & 47.3 & \textbf{45.3} & \textbf{92.8} & \textbf{92.1} & \textbf{92.4} \\
    \rowcolor{gray!20}
    \textbf{DeepResonance-$\beta$ (ours)} & \textbf{44.7} & \textbf{18.5} & \textbf{50.4} & \textbf{47.9} & \textbf{47.0} & \textbf{93.1} & \textbf{92.2} & \textbf{92.6} \\
    \hline
    \textit{MusiLingo (supervised performance upper bound)} & \textit{47.0} & -- & -- & -- & \textit{51.4} & -- & -- & \textit{92.9} \\
    \bottomrule
    \end{tabular}
    }
    \caption{\textbf{Results on MusicInstruct-short.} The top two performances are highlighted in \textbf{bold}.}
    \label{tab:musicinstructshort}
\end{table*}

\begin{table*}[t!]
    \centering
    \resizebox{\linewidth}{!}{
    \begin{tabular}{lrrrrrrrr}
    \toprule
    \textbf{Model} & \textbf{BLEU-1} & \textbf{BLEU} & \textbf{ROUGE-P} & \textbf{ROUGE-R} & \textbf{ROUGE-F1} & \textbf{BERT-P} & \textbf{BERT-R} & \textbf{BERT-F1} \\
    \toprule
    \textbf{NExT-GPT}~\cite{DBLP:conf/icml/Wu0Q0C24} & \textbf{25.6} & \textbf{5.1} & 22.6 & \textbf{31.2} & 23.7 & 87.6 & \textbf{88.4} & 88.0 \\
    \textbf{NExT-GPT w/ M$^2$UGen} & 18.9 & 3.3 & \textbf{38.4} & 19.7 & 24.2 & 89.6 & 87.2 & 88.4 \\
    \textbf{NExT-GPT w/ Music4way} & 20.6 & 3.0 & 36.7 & 20.4 & 25.0 & 89.7 & 87.7 & 88.7 \\
    \rowcolor{gray!20}
    \textbf{DeepResonance-$\alpha$ (ours)} & 22.6 & 3.7 & 35.4 & 21.6 & \textbf{25.3} & \textbf{89.9} & 87.9 & \textbf{88.8} \\
    \rowcolor{gray!20}
    \textbf{DeepResonance-$\beta$ (ours)} & \textbf{23.5} & \textbf{4.7} & \textbf{37.8} & \textbf{23.0} & \textbf{27.2} & \textbf{90.4} & \textbf{88.2} & \textbf{89.3} \\
    \hline
    \textit{OpenMU (supervised performance upper bound)} & \textit{45.7} & \textit{19.2} & \textit{36.7} & \textit{46.3} & \textit{39.8} & \textit{91.2} & \textit{92.4} & \textit{91.8} \\
    \bottomrule
    \end{tabular}
    }
    \caption{\textbf{Results on MTG.} The top two performances are highlighted in \textbf{bold}.}
    \label{tab:mtg}
\end{table*}

\begin{table*}[t]
    \centering
    \resizebox{\linewidth}{!}{
    \begin{tabular}{lrrrrrrrr}
    \toprule
    \textbf{Model} & \textbf{BLEU-1} & \textbf{BLEU} & \textbf{ROUGE-P} & \textbf{ROUGE-R} & \textbf{ROUGE-F1} & \textbf{BERT-P} & \textbf{BERT-R} & \textbf{BERT-F1} \\
    \toprule
    (1): \textbf{NExT-GPT w/ Music4way} & 34.1 & 13.8 & 48.2 & 37.9 & 39.2 & 92.1 & 90.5 & 91.3 \\
    (1) + MIE & 35.0 & 14.5 & 50.1 & 38.3 & 40.4 & 92.3 & 90.6 & 91.4 \\
    (1) + PT (6-layer) & \textbf{35.8} & \textbf{15.3} & 49.0 & \textbf{39.2} & 40.7 & 92.1 & 90.6 & 91.3 \\
    (1) + MIE + PT (6-layer) & 35.5 & 14.8 & 46.4 & 39.0 & 39.4 & 91.4 & 90.5 & 90.9 \\
    (1) + PT (2-layer) & 34.5 & 14.8 & 50.9 & 38.3 & 40.5 & 92.4 & 90.6 & 91.5 \\
    (1) + MIE + PT (2-layer) & 34.3 & 14.3 & 50.2 & 38.0 & 40.1 & 92.4 & 90.6 & 91.4 \\
    (1) + PT (1-layer) & 34.9 & 14.6 & 49.6 & 38.3 & 40.2 & 92.3 & \textbf{90.8} & 91.5 \\
    (1) + MIE + PT (1-layer) & \textbf{35.6} & 15.3 & 50.6 & \textbf{39.5} & \textbf{41.1} & 92.4 & \textbf{90.8} & 91.5 \\
    \hline
    (2): (1) + Music4way-MI2T \& Music4way-MV2T & 35.0 & 14.5 & 49.5 & 38.5 & 40.2 & 92.3 & 90.7 & 91.5 \\
    (2) + MIE (\textbf{DeepResonance-$\alpha$}) & 35.1 & 15.1 & \textbf{51.0} & 38.6 & 40.8 & \textbf{92.5} & 90.7 & \textbf{91.6} \\
    (2) + PT (6-layer) & 34.4 & 14.8 & 49.0 & 38.0 & 40.0 & 92.2 & 90.4 & 91.3 \\
    (2) + MIE + PT (6-layer) & 35.3 & 14.6 & 46.6 & 38.7 & 39.4 & 91.6 & 90.4 & 91.0 \\
    (2) + PT (2-layer) & 34.2 & 14.7 & \textbf{51.0} & 38.0 & 40.4 & \textbf{92.5} & 90.6 & 91.5 \\
    (2) + MIE + PT (2-layer) & 34.5 & 14.6 & 50.8 & 38.1 & 40.4 & \textbf{92.5} & 90.6 & 91.5 \\
    (2) + PT (1-layer) & 35.5 & 15.0 & 50.3 & \textbf{39.2} & 40.9 & 92.4 & \textbf{90.8} & \textbf{91.6} \\
    (2) + MIE + PT (1-layer) (\textbf{DeepResonance-$\beta$}) & \textbf{35.6} & \textbf{15.3} & \textbf{51.3} & 39.0 & \textbf{41.1} & \textbf{92.5} & \textbf{90.8} & \textbf{91.6} \\
    \hline
    \textbf{DeepResonance-$\alpha$} w/o v.c. & 35.0 & 14.7 & 49.8 & 38.5 & 40.4 & 92.3 & 90.6 & 91.4 \\
    \textbf{DeepResonance-$\alpha$} w/o m.f. & 34.9 & 14.5 & 49.8 & 38.5 & 40.3 & 92.3 & 90.7 & 91.5 \\
    \textbf{DeepResonance-$\beta$} w/o v.c. & 35.2 & 15.1 & 50.4 & 38.9 & 40.7 & 92.4 & 90.7 & 91.5 \\
    \textbf{DeepResonance-$\beta$} w/o m.f. & 35.2 & 14.7 & 49.3 & 38.7 & 40.2 & 92.2 & 90.7 & 91.4 \\
    \bottomrule
    \end{tabular}
    }
    \caption{\textbf{Ablation study on MusicQA.} The top two performances are highlighted in \textbf{bold}. ``v.c.'' and ``m.f.'' refer to visual captions and low-level musical features, respectively.}
    \label{tab:ablation1}
\end{table*}

\begin{table*}[t]
    \centering
    \resizebox{\linewidth}{!}{
    \begin{tabular}{lrrrrrrrr}
    \toprule
    \textbf{Model} & \textbf{BLEU-1} & \textbf{BLEU} & \textbf{ROUGE-P} & \textbf{ROUGE-R} & \textbf{ROUGE-F1} & \textbf{BERT-P} & \textbf{BERT-R} & \textbf{BERT-F1} \\
    \toprule
    (1): \textbf{NExT-GPT w/ Music4way} & 25.0 & 2.7 & 23.0 & 20.6 & 21.0 & \textbf{87.6} & 87.0 & 87.2 \\
    (1) + MIE & 25.4 & 2.6 & 22.8 & 21.3 & 21.1 & 87.5 & 87.0 & 87.2 \\
    (1) + PT (6-layer) & 24.3 & 2.5 & 21.3 & 20.5 & 20.0 & 87.1 & 86.6 & 86.8 \\
    (1) + MIE + PT (6-layer) & 21.9 & 1.7 & 21.1 & 18.5 & 18.6 & 86.9 & 86.1 & 86.5 \\
    (1) + PT (2-layer) & 24.7 & 2.4 & 21.8 & 20.6 & 20.2 & 87.3 & 86.9 & 87.1 \\
    (1) + MIE + PT (2-layer) & 25.5 & 2.5 & 22.2 & 21.2 & 20.8 & 87.4 & 87.0 & 87.2 \\
    (1) + PT (1-layer) & 25.7 & \textbf{2.8} & 22.7 & 21.4 & 21.1 & 87.5 & \textbf{87.1} & \textbf{87.3} \\
    (1) + MIE + PT (1-layer) & 25.8 & \textbf{2.8} & \textbf{23.5} & 21.5 & \textbf{21.6} & 87.5 & 87.0 & \textbf{87.3} \\
    \hline
    (2): (1) + Music4way-MI2T \& Music4way-MV2T & 25.9 & \textbf{2.8} & 23.1 & \textbf{21.6} & 21.4 & \textbf{87.6} & 87.0 & \textbf{87.3} \\
    (2) + MIE (\textbf{DeepResonance-$\alpha$}) & \textbf{26.0} & \textbf{3.0} & 23.4 & \textbf{21.8} & \textbf{21.6} & \textbf{87.6} & \textbf{87.1} & \textbf{87.3} \\
    (2) + PT (6-layer) & 21.6 & 1.7 & 21.3 & 18.0 & 18.5 & 86.9 & 86.1 & 86.5 \\
    (2) + MIE + PT (6-layer) & 21.6 & 1.8 & 21.2 & 18.2 & 18.5 & 87.0 & 86.1 & 86.5 \\
    (2) + PT (2-layer) & 24.9 & 2.5 & 22.0 & 20.8 & 20.4 & 87.3 & 86.9 & 87.1 \\
    (2) + MIE + PT (2-layer) & 25.1 & 2.4 & 22.1 & 21.0 & 20.5 & 87.3 & 86.9 & 87.1 \\
    (2) + PT (1-layer) & \textbf{25.9} & \textbf{2.8} & \textbf{23.5} & 21.4 & \textbf{21.6} & 87.5 & 87.0 & \textbf{87.3} \\
    (2) + MIE + PT (1-layer) (\textbf{DeepResonance-$\beta$}) & 25.8 & \textbf{2.8} & \textbf{23.6} & 21.4 & \textbf{21.6} & 87.5 & \textbf{87.1} & \textbf{87.3} \\
    \hline
    \textbf{DeepResonance-$\alpha$} w/o v.c. & 25.5 & 2.6 &  23.0 & 21.4 & 21.2 & 87.6 & 86.9 & 87.2 \\
    \textbf{DeepResonance-$\alpha$} w/o m.f. & \textbf{25.9} & 2.7 &  22.9 & \textbf{21.6} & 21.3 & \textbf{87.6} & 87.0 & \textbf{87.3} \\
    \textbf{DeepResonance-$\beta$} w/o v.c. & 25.8 & 2.7 &  22.8 & 21.5 & 21.2 & \textbf{87.6} & \textbf{87.1} & \textbf{87.3} \\
    \textbf{DeepResonance-$\beta$} w/o m.f. & 25.8 & 2.7 &  22.7 & 21.5 & 21.1 & 87.5 & \textbf{87.1} & \textbf{87.3} \\
    \bottomrule
    \end{tabular}
    }
    \caption{\textbf{Ablation study on MusicCaps.} The top two performances are highlighted in \textbf{bold}. ``v.c.'' and ``m.f.'' refer to visual captions and low-level musical features, respectively.}
    \label{tab:ablation2}
\end{table*}

\begin{table*}[t]
    \centering
    \resizebox{\linewidth}{!}{
    \begin{tabular}{lrrrrrrrr}
    \toprule
    \textbf{Model} & \textbf{BLEU-1} & \textbf{BLEU} & \textbf{ROUGE-P} & \textbf{ROUGE-R} & \textbf{ROUGE-F1} & \textbf{BERT-P} & \textbf{BERT-R} & \textbf{BERT-F1} \\
    \toprule
    (1): \textbf{NExT-GPT w/ Music4way} & 39.1 & 18.4 & 55.5 & 46.9 & 46.8 & 91.7 & 92.5 & 93.0 \\
    (1) + MIE & 39.5 & 18.2 & 55.8 & 46.5 & 46.8 & 93.7 & 92.5 & 93.0 \\
    (1) + PT (6-layer) & 35.9 & 14.5 & 50.2 & 43.2 & 42.7 & 92.8 & 91.9 & 92.3 \\
    (1) + MIE + PT (6-layer) & 33.4 & 14.1 & \textbf{58.0} & 40.5 & 45.0 & 93.6 & 91.0 & 92.3 \\
    (1) + PT (2-layer) & 39.6 & 18.7 & 56.7 & 46.3 & 47.3 & 93.9 & 92.4 & 93.1 \\
    (1) + MIE + PT (2-layer) & 39.9 & 19.1 & 57.2 & 46.9 & 47.7 & 93.9 & 92.5 & \textbf{93.2} \\
    (1) + PT (1-layer) & 40.3 & 19.3 & 57.6 & 46.8 & 47.9 & \textbf{94.0} & 92.5 & \textbf{93.2} \\
    (1) + MIE + PT (1-layer) & 39.9 & 19.1 & 56.7 & 47.1 & 47.5 & 93.8 & 92.5 & 93.1 \\
    \hline
    (2): (1) + Music4way-MI2T \& Music4way-MV2T & \textbf{40.7} & \textbf{19.8} & 57.0 & \textbf{48.0} & \textbf{48.2} & 93.9 & \textbf{92.6} & \textbf{93.2} \\
    (2) + MIE (\textbf{DeepResonance-$\alpha$}) & \textbf{40.9} & \textbf{19.9} & 57.8 & 47.5 & \textbf{48.4} & \textbf{94.0} & \textbf{92.6} & \textbf{93.3} \\
    (2) + PT (6-layer) & 33.5 & 14.2 & \textbf{58.0} & 40.4 & 45.0 & 93.6 & 91.0 & 92.3 \\
    (2) + MIE + PT (6-layer) & 33.6 & 14.2 & \textbf{58.0} & 40.5 & 45.1 & 93.6 & 91.0 & 92.3 \\
    (2) + PT (2-layer) & 39.6 & 18.7 & 56.7 & 46.3 & 47.3 & 93.9 & 92.4 & 93.1 \\
    (2) + MIE + PT (2-layer) & 39.2 & 18.8 & 57.1 & 46.8 & 47.4 & 93.9 & 92.4 & 93.1 \\
    (2) + PT (1-layer) & 40.3 & 19.5 & 57.3 & 47.4 & 48.0 & 93.9 & 92.5 & \textbf{93.2} \\
    (2) + MIE + PT (1-layer) (\textbf{DeepResonance-$\beta$}) & 39.9 & 19.3 & 57.3 & 47.3 & 47.8 & 93.9 & 92.5 & \textbf{93.2} \\
    \hline
    \textbf{DeepResonance-$\alpha$} w/o v.c. & \textbf{40.7} & 19.7 & 57.2 & 47.7 & 48.1 & \textbf{94.0} & \textbf{92.6} & \textbf{93.2} \\
    \textbf{DeepResonance-$\alpha$} w/o m.f. & 40.5 & 19.6 & 56.7 & \textbf{48.1} & 48.0 & 93.9 & \textbf{92.6} & \textbf{93.2} \\
    \textbf{DeepResonance-$\beta$} w/o v.c. & 39.5 & 18.5 & 55.0 & 47.2 & 46.7 & 93.6 & 92.5 & 93.0 \\
    \textbf{DeepResonance-$\beta$} w/o m.f. & 40.0 & 19.1 & 56.3 & 47.7 & 47.5 & 93.8 & 92.5 & 93.1 \\
    \bottomrule
    \end{tabular}
    }
    \caption{\textbf{Ablation study on Music4way-MusicCaps.} The top two performances are highlighted in \textbf{bold}. ``v.c.'' and ``m.f.'' refer to visual captions and low-level musical features, respectively.}
    \label{tab:ablation3}
\end{table*}

\begin{table*}[t]
    \centering
    \resizebox{\linewidth}{!}{
    \begin{tabular}{lrrrrrrrr}
    \toprule
    \textbf{Model} & \textbf{BLEU-1} & \textbf{BLEU} & \textbf{ROUGE-P} & \textbf{ROUGE-R} & \textbf{ROUGE-F1} & \textbf{BERT-P} & \textbf{BERT-R} & \textbf{BERT-F1} \\
    \toprule
    (1): \textbf{NExT-GPT w/ Music4way} & 24.2 & 1.7 & 25.1 & 23.6 & 22.0 & 85.7 & 85.3 & 85.4 \\
    (1) + MIE & 23.7 & 1.7 & 30.0 & 18.5 & 22.1 & 87.9 & 85.8 & 86.8 \\
    (1) + PT (6-layer) & 20.6 & 1.3 & 30.8 & 17.8 & 21.0 & 87.6 & 85.5 & 86.5 \\
    (1) + MIE + PT (6-layer) & 12.0 & 0.3 & 25.4 & 14.6 & 16.2 & 85.0 & 84.1 & 84.5 \\
    (1) + PT (2-layer) & 23.0 & 1.8 & 32.8 & 19.9 & 23.3 & 87.8 & 85.9 & 86.8 \\
    (1) + MIE + PT (2-layer) & 10.7 & 0.2 & 29.2 & 11.8 & 16.0 & 86.9 & 84.3 & 85.6 \\
    (1) + PT (1-layer) & 19.4 & 1.4 & 30.5 & 17.1 & 20.7 & 87.6 & 85.5 & 86.6 \\
    (1) + MIE + PT (1-layer) & 15.1 & 0.6 & 30.1 & 14.6 & 18.8 & 87.4 & 84.9 & 86.1 \\
    \hline
    (2): (1) + Music4way-MI2T \& Music4way-MV2T & 43.8 & 8.7 & 28.0 & 34.0 & 30.1 & 88.8 & 89.5 & 89.1 \\
    (2) + MIE (\textbf{DeepResonance-$\alpha$}) & 48.7 & 16.4 & \textbf{37.1} & 36.8 & 36.2 & \textbf{90.3} & \textbf{90.2} & \textbf{90.2} \\
    (2) + PT (6-layer) & 48.6 & 15.5 & \textbf{37.1} & 36.2 & 36.1 & 90.0 & 90.1 & 90.0 \\
    (2) + MIE + PT (6-layer) & \textbf{49.1} & 15.7 & 36.3 & 36.9 & 36.1 & 90.1 & 90.1 & 90.1 \\
    (2) + PT (2-layer) & 47.9 & 15.9 & 36.7 & 36.5 & 35.8 & 90.1 & 90.1 & 90.1 \\
    (2) + MIE + PT (2-layer) & 49.0 & \textbf{17.1} & 36.7 & \textbf{38.3} & \textbf{36.7} & 90.0 & \textbf{90.3} & 90.1 \\
    (2) + PT (1-layer) & 48.4 & 16.6 & \textbf{37.3} & 37.0 & 36.4 & \textbf{90.3} & \textbf{90.2} & \textbf{90.2} \\
    (2) + MIE + PT (1-layer) (\textbf{DeepResonance-$\beta$}) & \textbf{49.2} & \textbf{17.2} & 36.8 & \textbf{38.3} & \textbf{36.8} & 90.1 & \textbf{90.3} & \textbf{90.2} \\
    \bottomrule
    \end{tabular}
    }
    \caption{\textbf{Ablation study on Music4way-MI2T.} The top two performances are highlighted in \textbf{bold}.}
    \label{tab:ablation4}
\end{table*}

\begin{table*}[t]
    \centering
    \resizebox{\linewidth}{!}{
    \begin{tabular}{lrrrrrrrr}
    \toprule
    \textbf{Model} & \textbf{BLEU-1} & \textbf{BLEU} & \textbf{ROUGE-P} & \textbf{ROUGE-R} & \textbf{ROUGE-F1} & \textbf{BERT-P} & \textbf{BERT-R} & \textbf{BERT-F1} \\
    \toprule
    (1): \textbf{NExT-GPT w/ Music4way} & 25.0 & 1.8 & 25.4 & 24.2 & 22.5 & 85.7 & 85.4 & 85.5 \\
    (1) + MIE & 14.2 & 0.6 & 34.7 & 14.9 & 20.1 & 88.1 & 84.9 & 86.4 \\
    (1) + PT (6-layer) & 17.0 & 1.1 & 34.6 & 16.6 & 20.8 & 88.1 & 85.3 & 86.7 \\
    (1) + MIE + PT (6-layer) & 16.3 & 0.7 & 28.3 & 16.1 & 19.0 & 86.2 & 84.9 & 85.5 \\
    (1) + PT (2-layer) & 23.3 & 1.8 & 31.9 & 20.6 & 23.2 & 87.6 & 85.8 & 86.7 \\
    (1) + MIE + PT (2-layer) & 6.4 & 0.2 & 36.2 & 10.3 & 15.2 & 87.5 & 83.8 & 85.6 \\
    (1) + PT (1-layer) & 19.5 & 1.3 & 30.8 & 17.0 & 20.8 & 87.7 & 85.5 & 86.6 \\
    (1) + MIE + PT (1-layer) & 10.3 & 0.6 & \textbf{37.5} & 13.5 & 18.9 & 88.3 & 84.7 & 86.5 \\
    \hline
    (2): (1) + Music4way-MI2T \& Music4way-MV2T & 43.0 & 8.8 & 28.0 & 32.7 & 29.6 & 88.8 & 89.4 & 89.1 \\
    (2) + MIE (\textbf{DeepResonance-$\alpha$}) & 48.9 & 16.6 & \textbf{37.4} & 37.0 & 36.5 & \textbf{90.3} & 90.2 & \textbf{90.2} \\
    (2) + PT (6-layer) & 48.5 & 15.4 & 37.0 & 36.2 & 36.1 & 89.9 & 90.1 & 90.0 \\
    (2) + MIE + PT (6-layer) & 48.7 & 15.3 & 36.9 & 36.2 & 35.9 & \textbf{90.4} & 90.0 & \textbf{90.2} \\
    (2) + PT (2-layer) & \textbf{49.0} & 16.9 & 36.6 & 37.9 & 36.5 & 90.0 & \textbf{90.3} & 90.1 \\
    (2) + MIE + PT (2-layer) & 48.9 & \textbf{17.0} & 36.6 & \textbf{38.4} & \textbf{36.7} & 90.0 & \textbf{90.3} & 90.1 \\
    (2) + PT (1-layer) & 48.5 & 16.5 & 37.0 & 36.9 & 36.2 & 90.2 & 90.2 & \textbf{90.2} \\
    (2) + MIE + PT (1-layer) (\textbf{DeepResonance-$\beta$}) & \textbf{49.0} & \textbf{17.2} & 36.7 & \textbf{38.4} & \textbf{36.8} & \textbf{90.3} & \textbf{90.3} & \textbf{90.3} \\
    \bottomrule
    \end{tabular}
    }
    \caption{\textbf{Ablation study on Music4way-MV2T.} The top two performances are highlighted in \textbf{bold}.}
    \label{tab:ablation5}
\end{table*}

\begin{table*}[t]
    \centering
    \resizebox{\linewidth}{!}{
    \begin{tabular}{lrrrrrrrr}
    \toprule
    \textbf{Model} & \textbf{BLEU-1} & \textbf{BLEU} & \textbf{ROUGE-P} & \textbf{ROUGE-R} & \textbf{ROUGE-F1} & \textbf{BERT-P} & \textbf{BERT-R} & \textbf{BERT-F1} \\
    \toprule
    (1): \textbf{NExT-GPT w/ Music4way} & 29.4 & 2.8 & 35.9 & 22.7 & 27.0 & 89.7 & 87.1 & 88.4 \\
    (1) + MIE & 16.2 & 1.2 & 39.1 & 16.7 & 22.8 & 89.8 & 85.9 & 87.8 \\
    (1) + PT (6-layer) & 3.7 & 0.0 & 10.9 & 10.2 & 7.0 & 70.9 & 77.2 & 73.8 \\
    (1) + MIE + PT (6-layer) & 21.3 & 1.6 & 34.2 & 19.6 & 23.5 & 88.3 & 85.9 & 87.0 \\
    (1) + PT (2-layer) & 7.4 & 0.5 & 36.9 & 10.8 & 14.4 & 84.3 & 80.8 & 82.5 \\
    (1) + MIE + PT (2-layer) & 1.3 & 0.0 & \textbf{41.5} & 7.5 & 12.1 & 87.3 & 82.4 & 84.8 \\
    (1) + PT (1-layer) & 23.5 & 1.9 & 34.7 & 20.0 & 24.1 & 88.9 & 86.1 & 87.5 \\
    (1) + MIE + PT (1-layer) & 1.6 & 0.0 & \textbf{43.5} & 8.2 & 13.4 & 87.9 & 82.6 & 85.2 \\
    \hline
    (2): (1) + Music4way-MI2T \& Music4way-MV2T & \textbf{54.4} & \textbf{13.7} & 34.3 & \textbf{34.8} & \textbf{34.4} & \textbf{90.9} & \textbf{90.5} & \textbf{90.7} \\
    (2) + MIE (\textbf{DeepResonance-$\alpha$}) & \textbf{37.2} & \textbf{6.0} & 36.0 & \textbf{25.7} & \textbf{29.6} & \textbf{90.8} & \textbf{88.3} & \textbf{89.5} \\
    (2) + PT (6-layer) & 36.9 & 4.6 & 33.2 & 25.2 & 28.2 & 90.1 & 87.8 & 88.9 \\
    (2) + MIE + PT (6-layer) & 34.1 & 4.1 & 34.2 & 24.3 & 27.9 & 90.2 & 87.6 & 88.9 \\
    (2) + PT (2-layer) & 17.8 & 1.5 & 36.2 & 17.1 & 20.3 & 88.0 & 84.9 & 86.3 \\
    (2) + MIE + PT (2-layer) & 36.2 & 3.4 & 28.6 & 25.0 & 25.9 & 88.6 & 87.3 & 87.9 \\
    (2) + PT (1-layer) & 24.7 & 2.4 & 34.0 & 20.3 & 23.5 & 88.6 & 86.0 & 87.2 \\
    (2) + MIE + PT (1-layer) (\textbf{DeepResonance-$\beta$}) & 33.5 & 3.9 & 34.0 & 23.9 & 27.4 & 90.0 & 87.5 & 88.7 \\
    \bottomrule
    \end{tabular}
    }
    \caption{\textbf{Ablation study on Music4way-Any2T.} The top two performances are highlighted in \textbf{bold}.}
    \label{tab:ablation6}
\end{table*}

\end{document}